\documentclass[%
 reprint,
 amsmath,amssymb,
aps,
pra,
]{revtex4-1}

\usepackage{amsmath}
\usepackage{hyperref}
\usepackage{natbib}
\usepackage{graphicx}
\usepackage{dcolumn}
\usepackage{bm}
\usepackage{subfig}
\usepackage{color}

\DeclareMathOperator{\Tr}{Tr}

\begin{document}


\title{Stability analysis of a periodic system of relativistic current filaments}

\author{A. Vanthieghem}
 \email{arno.vanthieghem@iap.fr}
 \affiliation{Sorbonne Universit\'e, CNRS, UMR 7095, Institut d'Astrophysique de Paris,\\
 98 bis bd Arago, 75014 Paris, France\\}
\affiliation{Sorbonne Universit\'es, Institut Lagrange de Paris (ILP),\\98 bis bd Arago, 75014 Paris, France}

\author{M. Lemoine}
 \email{lemoine@iap.fr}
 \affiliation{Sorbonne Universit\'e, CNRS, UMR 7095,\\
 Institut d'Astrophysique de Paris,
 98 bis bd Arago, 75014 Paris, France}

\author{L. Gremillet}
\email{laurent.gremillet@cea.fr}
\affiliation{CEA, DAM, DIF, F-91297 Arpajon, France}


\date{\today}

\begin{abstract}
The nonlinear evolution of current filaments generated by the Weibel-type filamentation instability is a topic of prime interest
in space and laboratory plasma physics. In this paper, we investigate the stability of a stationary periodic chain of nonlinear current filaments in
counterstreaming pair plasmas. We make use of a relativistic four-fluid model and apply the Floquet theory to compute the two-dimensional
unstable eigenmodes of the spatially periodic system. We examine three different cases, characterized by various levels of nonlinearity and
asymmetry between the plasma streams: a weakly nonlinear symmetric system, prone to purely transverse merging modes; a strongly nonlinear
symmetric system, dominated by coherent drift-kink modes whose transverse periodicity is equal to, or an integer fraction of the unperturbed filaments;
a moderately nonlinear asymmetric system, subject to a mix of kink and bunching-type perturbations. The growth rates and profiles of the numerically
computed  eigenmodes agree with particle-in-cell simulation results. In addition, we derive an analytic criterion for the transition between dominant
filament-merging and drift-kink instabilites in symmetric two-beam systems.
 
\end{abstract}

\maketitle

\section{\label{sec:level1_intro}Introduction}

The current filamentation instability (CFI) emerges as an essential process in various fields of plasma physics. 
It develops in anisotropic plasmas (where it is usually referred to as the Weibel instability) \cite{Weibel_1959, Davidson_1972}
or multi-stream plasmas \cite{Fried_1959, Califano_1997}, giving rise to kinetic-scale current filaments of alternating sign,
normal to the direction of larger temperature or drift. The associated electromagnetic fluctuations can cause efficient particle
scattering and deceleration \cite{Lee_1973, Honda_2000, Gedalin_2008}, which makes the CFI a likely key player in the formation of
astrophysical collisionless shocks \cite{Moiseev_1963, Medvedev_1999, Lemoine_2011, Pelletier_2017}. There, the CFI arises from
the interaction of a beam of energized particles issued from a central engine (or reflected off the shock front) with the ambient (or upstream) plasma. First-principles kinetic simulations
\cite{Frederiksen_2004, Kato_2007, Spitkovsky_2008a, Keshet_2008, Martins_2009, Nishikawa_2009, Sironi_2013, Ardaneh_2015} show that an electromagnetic
barrier then develops, which dissipates the upstream kinetic energy and promotes Fermi acceleration of the suprathermal particles needed to
sustain the CFI upstream of the shock. Moreover, the synchrotron emission of the energized particles in the downstream magnetic turbulence is
believed to account for the broadband photon spectra of various powerful astrophysical sources \cite{Sironi_2009a, Frederiksen_2010, Keenan_2013}.
The Weibel instability also appears to be influential in magnetic reconnection physics, by controling the current layer dynamics in
magnetized electron-positron ($e^-e^+$) pair plasmas \cite{Swisdak_2008}.

On the laboratory side, the CFI plays a major role in particle beam \cite{Molvig_1975, Lee_1983, Allen_2012} and laser-plasma
experiments, be it in the context of inertial confinement fusion \cite{Ramani_1978, True_1985, Epperlein_1986, Masson-Laborde_2010}
or relativistic laser-plasma interactions \cite{Ruhl_2002, Sentoku_2003, Tatarakis_2003, Wei_2004, Adam_2006, Debayle_2010, Mondal_2012, Quinn_2012}.
In the latter case, copious numbers of fast electrons are driven by the laser pulse into the bulk plasma, where they fragment into small-scale
magnetic filaments. The scatterings undergone by the fast electrons can lead to large angular divergence, and hence hamper applications involving
high electron flux densities, such as the fast ignition scheme \cite{Tabak_1994, Robinson_2014} or ion acceleration \cite{Gode_2017, Scott_2017}.
On the other hand, the CFI can be purposefully triggered in high-energy laser-driven plasma collisions \cite{Drake_2012, Fox_2013, Huntington_2015, Ruyer_2016}
or relativistic-intensity laser-plasma interactions \cite{Fiuza_2012, Ruyer_2015b, Sarri_2015, Lobet_2015, Warwick_2017}, designed as
testbeds for astrophysical shock models.


In past decades, the linear theory of the CFI has been extensively studied in various parameter ranges, using fluid or kinetic approaches,  whether relativistic or
not \cite{Davidson_1972, Yang_1993, Califano_1997, Silva_2002, Schaefer-Rolffs_2006, Yoon_2007, Bret_2007, Achterberg_2007_I, Cottrill_2008, Hao_2008a, Bret_2010a}.
Its nonlinear evolution has been mostly investigated using particle-in-cell (PIC) kinetic simulations, showing that the early-time instability growth is generally
arrested by magnetic trapping inside the filaments \cite{Davidson_1972, Wallace_1987, Yang_1994, Califano_1998a, Kato_2005, Okada_2007, Kaang_2009, Bret_2013},
or quasilinear heating in the weak-growth limit \cite{Park_2010, Pokhotelov_2011}. The evolution of the ensemble of magnetic filaments that are formed
following this primary saturation is, however, still an open problem. PIC simulations performed in the plane normal to the plasma flow indicate that the instability dynamics
becomes governed by successive coalescences between imperfectly screened filaments \cite{Lee_1973, Honda_2000, Sakai_2002, Silva_2003, Jaroschek_2005}.
A number of analytic models, often heuristic, have been proposed to describe this essentially transverse, filament-merging instability (FMI)
\cite{Honda_2000, Medvedev_2005, Achterberg_2007_II, Polomarov_2008, Stockem-Novo_2015, Ruyer_2015a}. By contrast, 3D or 2D simulations that resolve
the plasma flow axis show that modes with nonzero longitudinal wavenumber can alter the filament dynamics and mergers. Such modes have been interpreted
either as variants of the oblique waves \cite{Bret_2010a} that develop in homogeneous two-stream systems \cite{Silva_2003, Jaroschek_2005}, or as drift
kink instabilities (DKI) \cite{Daughton_1998, Karimabadi_2003a, Zenitani_2007, Ruyer_2016b, Barkov_2016} arising in isolated current filaments (analogous to current sheets in 2D) \cite{Milosavljevic_2006a,Ruyer_2018}.
Kelvin-Helmholtz-type modes have also been predicted to grow in the velocity shear layer between neighboring filaments \cite{Das_2001, Jain_2003}. 

Further understanding of the nonlinear evolution of the CFI therefore involves a comprehensive modeling of the unstable dynamics of an ensemble of
self-pinched filaments. Clearly, all previous works failed to provide such a unified picture, since they neglected either the nonlinearity (inhomogeneity) of
the filaments or the collective couplings between them. Here, by contrast, we present an exact two-dimensional (2D) stability analysis of a periodic chain of relativistic current
filaments, which treats on an equal footing all FMI and DKI-type modes. The unperturbed equilibrium system consists of a transverse stationary electromagnetic
wave embedded in two counterstreaming, neutral $e^-e^+$ pair flows. For the sake of simplicity, we make use of a warm-fluid model for the four plasma components
at play. The equilibrium system is then described by two coupled sinh-Gordon-type equations for the electromagnetic field potentials. We then exploit the periodicity of
the system in the transverse direction to extract its eigenmodes using the Floquet theory. A standard technique is to decompose the solution in a Fourier series, and to
solve the resultant infinite Hill's determinant \cite{Bertrand_1971,Quesnel_1997,Kuo_1997,Faith_1997}. Here, instead, we employ another method, based on a fundamental
theorem in the Floquet theory, which gives the full set of eigenmodes without having to derive Hill's determinant \cite{Romeiras_1986,Hada_1992,Zhang_2012}.
Our analysis only addresses the eigenmodes developing in the in-flow plane, which have been found to mainly control the long-term CFI evolution \cite{Sironi_2009b}.

The structure of the paper is as follows. In Sect.~\ref{sec:level1_model}, after presenting the theoretical framework, we derive the equilibrium equations,
and analytically solve them for two symmetric counterstreaming pair flows. In passing, we show that the well-known Harris solution is locally recovered
in the limit of strongly pinched filaments. We then derive the set of linearized perturbation equations, and in Sect.~\ref{sec:level1_Floquet}, we
detail the numerical method used to compute the unstable eigenmodes. The numerical solver is validated in the homogeneous limit through comparison
with the solutions of the standard relation dispersion. We then study three different configurations depending on the level  of nonlinearity and asymmetry
of the four-fluid system. In Sect.~\ref{sec:level1_weak}, we consider the case of a weakly nonlinear, symmetric system: its purely transverse, FMI-type
eigenmodes are solved, and found in good agreement with the results of a PIC simulation. Section~\ref{sec:level1_strong} deals with the 2D eigenmodes
developing in a symmetric system of varying nonlinearity. We demonstrate that merging modes are superseded by DKI-type modes above a threshold
nonlinearity level, which we evaluate analytically. The dominant DKI modes, of same transverse periodicity as the equilibrium system, turn out to be
very similar to those developing in an isolated current sheet. Again, the theoretical results are successfully confronted to PIC simulations.
Finally, we briefly examine the case of a moderately nonlinear asymmetric system in Sec.~\ref{sec:level1_asym}. Our illustrative calculation, supported by
PIC simulations, predicts a dominant mode which consists of coherent kink and bunching-type perturbations, nonlinearly evolving into an obliquely striped pattern.

\section{\label{sec:level1_model}Modeling a periodic chain of nonlinear current filaments}

We consider a 2D $(x,y)$ plasma comprising two counterstreaming, neutral pair beams, drifting along the $x$-axis. There is no external magnetic field,
so that the electric and magnetic fields, $\mathbf{E}=(E_x,E_y,0)$ and $\mathbf{B}=(0,0,B_z)$, are self-consistently generated by the plasma current
and charge modulations. In the equilibrium unperturbed state, the pair beams are periodically modulated along the tranverse $y$-axis, thus forming a
chain of current filaments of alternating sign. The equilibrium results from a balance between the Lorentz and thermal pressure forces inside the filaments.  

\subsection{\label{sec:level2_basic}Basic equations}

We now present the fundamental equations used to describe the unstable evolution of the initially modulated four-fluid pair plasma. Unless otherwise noted,
we use cgs Gaussian units. Each species $\alpha$ is characterized by its charge ($q_\alpha \equiv Z_\alpha e$, where $e$ is the elementary charge and $Z_\alpha = \pm 1$), 
drift velocity ($\mathbf{v}_\alpha = \bm{\beta}_\alpha c$, where $c$ is the velocity of light), Lorentz factor ($\gamma_\alpha \equiv 1/\sqrt{1-\beta_\alpha^2}$),
pressure ($p_\alpha$), rest-frame density ($n_\alpha$), lab-frame density ($d_\alpha \equiv \gamma_\alpha n_\alpha$), and polytropic index ($\Gamma_\mathrm{ad, \alpha}$).

The continuity and warm-fluid momentum equations for species $\alpha$ read
\begin{align}
 &\partial_{ct} d_\alpha + \bm{\nabla} \cdot (d_\alpha \bm{\beta}_\alpha) = 0 \, \label{eq:continuity} \,, \\
 &\gamma^2_\alpha  (p_\alpha + \epsilon_\alpha) (\partial_{ct} + \bm{\beta}_\alpha \cdot \bm{\nabla}) \bm{\beta}_\alpha =-\bm{\nabla} p_\alpha \nonumber\\
 &+ q_{\alpha} d_\alpha (\mathbf{E} + \bm{\beta}_\alpha \times \mathbf{B}) - \bm{\beta}_\alpha (q_\alpha d_\alpha \mathbf{E} \cdot \bm{\beta}_\alpha + 
 \partial_{ct} p_\alpha) \label{eq:momentum} \,,
\end{align}
where $\epsilon_\alpha = n_\alpha m c^2 + p_\alpha/(\Gamma_\mathrm{ad,\alpha} - 1)$ is the rest-frame energy density, $m$ is the electron mass,
and $\partial_{c t} \equiv c^{-1} \partial/\partial t$. The above equations should be solved along with the Maxwell equations:
\begin{align}
 &\bm{\nabla} \times \mathbf{B} = 4 \pi \sum\limits_\alpha  q_\alpha  d_\alpha \bm{\beta}_\alpha +  \partial_{ct} \mathbf{E} \,, \label{eq:Maxwell} \\
 &\bm{\nabla} \times \mathbf{E} = - \partial_{ct}  \mathbf{B} \,, \label{eq:Faraday} \\
 &\bm{\nabla} \cdot \mathbf{E} = 4 \pi \sum\limits_\alpha q_\alpha d_\alpha \label{eq:Gauss} \,.
\end{align} 
Note that Eq.~\eqref{eq:Gauss}, which is redundant with Eq.~\eqref{eq:continuity}, is exploited to ease the derivation of the stationary state
in the following section.

\subsection{\label{sec:level2_stat}Stationary state}

The stationary state is obtained by setting $\partial_{ct} = 0$ and $\partial_{x} = 0$ in the above equations. The unperturbed quantities are denoted by the index `0'.
From Eq.~\eqref{eq:momentum}, the equilibrium pressure of each fluid satisfies
\begin{equation}\label{eq:momentum_stat}
 \partial_y p_{\alpha 0} =  q_\alpha d_{\alpha 0}  \left(E_{0y} - \beta_{\alpha 0} B_{0z}\right) \,,
\end{equation}
while the Amp\`ere-Maxwell and Gauss-Maxwell equations become
\begin{align}
 &\partial_y B_{0z} =  4 \pi \sum\limits_\alpha q_\alpha d_{\alpha 0} \beta_{\alpha 0} \label{eq:Ampere_stat} \,, \\
 &\partial_y E_{0y} =  4 \pi \sum\limits_\alpha q_\alpha d_{\alpha 0}  \label{eq:Gauss_stat} \,.
\end{align}
where $\beta_{\alpha 0} \equiv \beta_{\alpha 0 x}$ is introduced to simplify notations.

The set of fluid equations is closed using the isothermal condition
\begin{equation}\label{eq:isothermal}
p_{\alpha 0} = n_{\alpha 0} T_\alpha m c^2,
\end{equation}
where $T_\alpha$ is the rest-frame temperature normalized to the electron rest mass energy.

To make analytical progress, it is convenient to use the potential equations:
\begin{flalign}
  \mathbf{B}_0 &= \bm{\nabla} \times \mathbf{A}_0 \,, \\
  \mathbf{E}_0 &= -\bm{\nabla} \phi_0 \,.
\end{flalign}
Combining these equations with Eqs.~\eqref{eq:isothermal} and \eqref{eq:momentum_stat} readily gives, after integration, the lab-frame density as
\begin{equation}\label{eq:density}
d_{\alpha 0} = N_\alpha \gamma_{\alpha 0} \exp \left[ -\frac{\gamma_{\alpha 0} q_\alpha} {T_\alpha m c^2} \left(\phi_0 - \beta_{\alpha 0} A_{0x}\right) \right] \,,
\end{equation}
where the factor $N_\alpha$ corresponds to the proper density in the (homogeneous) limit of vanishing fields ($A_{0x} \rightarrow 0$,
$\phi_0 \rightarrow 0$). The same relation would have been obtained in a kinetic framework using a J\"uttner-Synge distribution
\cite{Kocharovsky_2010}. 

In the following, the equilibrium electron and positron streams that make up a pair beam are assumed to share the same characteristics (except for
their opposite charge). Therefore, the unperturbed system will be defined by two sets of quantities $(n_{p0}, \gamma_{p0}, T_{p0}, \Gamma_\mathrm{ad,p})$
and $(n_{b 0}, \gamma_{b 0}, T_{b 0}, \Gamma_\mathrm{ad,b})$, where the labels $b$ and $p$ respectively stand for `beam' and `plasma'. In
Eqs.~\eqref{eq:Ampere_stat} and \eqref{eq:Gauss_stat}, the source terms can then be grouped pair-wise, leading to the following defining equations
for the stationary fluid-Maxwell system:
\begin{flalign}
 &\partial_y^2 A_{0x} =  4 \pi e \nonumber\\
 &\times \sum\limits_\alpha N_\alpha \gamma_{\alpha 0} \beta_{\alpha 0} \sinh\left[\frac{\gamma_{\alpha 0} e}{T_\alpha m c^2}
 \left(\phi_0 - \beta_{\alpha 0} A_{0x}\right)\right] \,, \label{eq:A_stat} \\
 &\partial_y^2 \phi_0 =  4 \pi e \sum\limits_\alpha N_\alpha \gamma_{\alpha 0} \sinh \left[\frac{\gamma_{\alpha 0} e}{T_\alpha m c^2}
 \left(\phi_0 - \beta_{\alpha 0} A_{0x}\right) \right] \label{eq:phi_stat} \,.
\end{flalign}
We seek a periodic solution for $A_{0x}(y)$ and $\phi_0(y)$ with a single maximum per (unknown) period $\lambda_0$. Introducing $a_0 \equiv (e/mc^2) \mathrm{max}_y A_{0x}$,
the boundary conditions are chosen to be $A_{0x}(0) = A_{0x}(\lambda_0) = (m c^2/e) a_0$, $\phi_0(0) = \phi_0(\lambda_0)$, and $\partial_y A_{0x}(0) = \partial_y \phi_0(0) = 0$.  

\begin{figure}[t!]
  \centering
  \includegraphics[width=0.8\hsize]{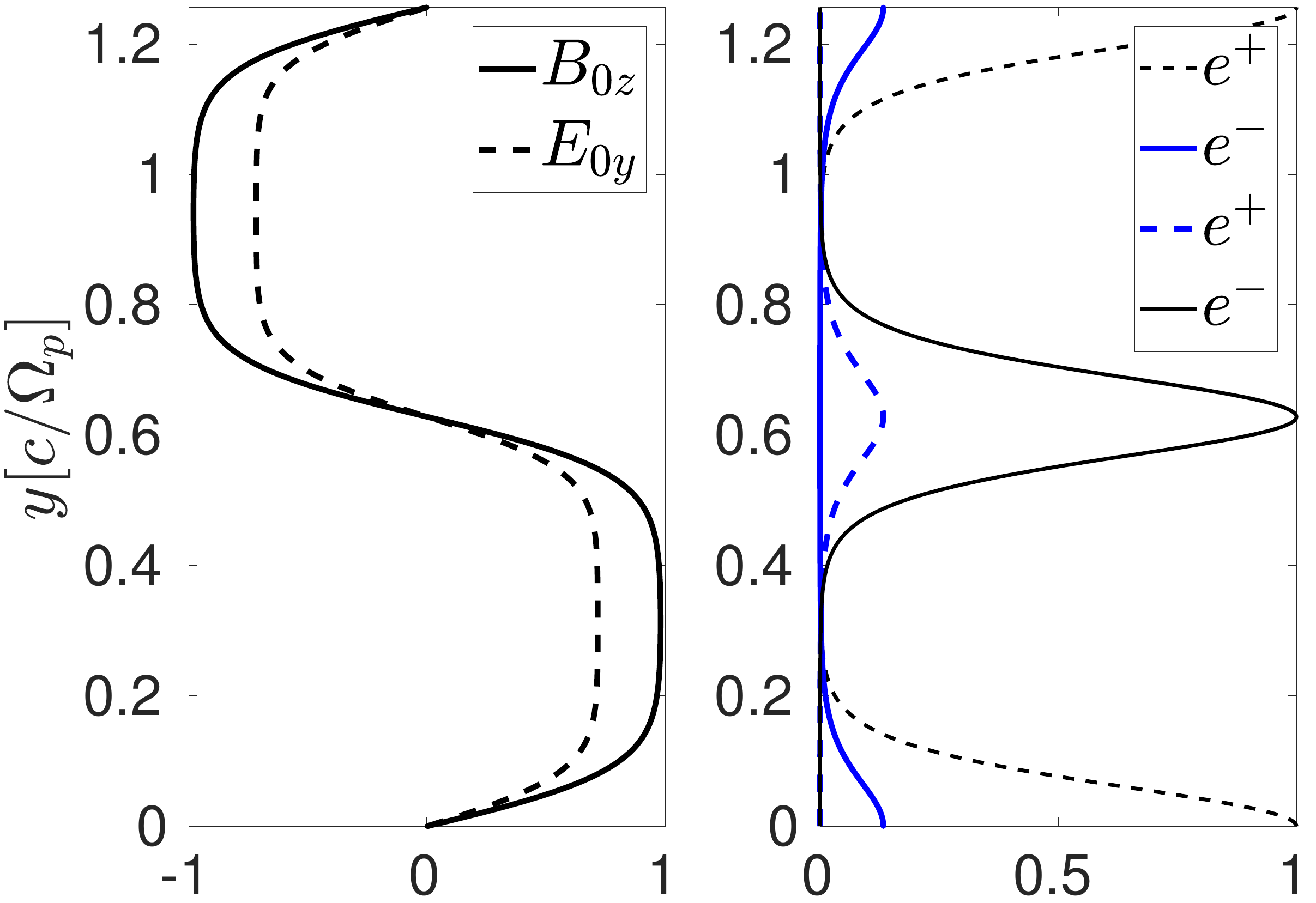}
  \caption{Stationary solution to the fluid-Maxwell equations in an $e^-e^+$ system composed of a hot beam ($T_{b0} = 1$, $\beta_{b0} = -0.995$)
  and a background cold plasma ($N_p/N_b=1$, $T_{p0}=0.1$, $\beta_{p0}=0.995$). The vector potential maximum is $a_0=0.25$, resulting in a
  wavelength $\lambda_0=1.26 \,c/\Omega_p$.
  Left panel: $B_{0z}(y)$ (solid line) and $E_{0y}(y)$ (dashed line), normalized to $m c \Omega_p/e$, where $\Omega_p$ is the relativistic frequency associated
  with the peak electron (or positron) frequency.
  Right panel: Density profiles of the electrons (solid lines) and positrons (dashed lines) in the hot (blue) and cold (black) beams,
  normalized to the maximum cold-beam electron (or positron) density.}
  \label{fgr:p_prof_strong}
\end{figure}

In the general case, the eigenvalue problem \eqref{eq:A_stat},\eqref{eq:phi_stat} must be solved numerically. An example is given in Fig.~\ref{fgr:p_prof_strong} where the profiles of $B_z$, $E_y$ and $d_{b,p}$ are plotted over one period in the case of a hot beam ($T_{b0} = 1$, $\beta_{b0} = -0.995$) streaming against a cold beam ($T_{p0}=0.1$, $\beta_{p0}=0.995$) with $N_p/N_b=1$. The vector potential maximum is set to $a_0 =0.25$, yielding $\lambda_0 \simeq 1.26 \,c/\Omega_p$, $\max_y B_{0z}\simeq 0.98\,m c \Omega_p/e$ and $\max_y E_{y0}\simeq 0.72 \,mc\Omega_p/e$, where
$\Omega_p$ is the rest-frame plasma frequency of the cold-beam electrons (or positrons),
\begin{equation}
\Omega_p = \sqrt{\frac{4\pi e^2 \mathrm{max}_y \left( n_{p0} \right)}{m}}.
\end{equation} 

Analytic expressions for the wavelength $\lambda_0$ can be derived for two symmetric pair beams ($| \beta_{b0} | =| \beta_{p0} |=\beta_0$,
$\gamma_{b0}=\gamma_{p0}=\gamma_0$, $T_{b0}=T_{p0}=T_0$, $N_b = N_p = N$), which can then serve as initial guesses for the
boundary value solver. Due to the symmetry, the electrostatic field vanishes. Setting $\phi_0 = 0$, the problem reduces to solving Eq.~\eqref{eq:A_stat}.
Let us first consider the weak-field limit, in which case we readily obtain
\begin{equation}
  \lambda_0 = \frac{\pi \sqrt{T_0}}{\beta_0 \gamma_0}\ \frac{c}{\omega_p} \,,
\end{equation}
where $\omega_p = \sqrt{4 \pi e^2 N/m}$ is the relativistic plasma frequency of a single (electron or positron) species, expressed in terms of
the density parameter $N$ appearing in Eq.~\eqref{eq:density}. 

In the more general, nonlinear symmetric case, the periodicity $\lambda_0$ depends on $a_0$ through the following exact analytic expression
\begin{equation}\label{eq:lambda}
  \lambda_0 = \frac{2 \sqrt{T_0}}{\beta_0 \gamma_0} K \left[-\sinh^2 \left(\frac{\xi}{2} \right) \right]
  \frac{c}{\omega_p} \,,
\end{equation}
with the nonlinearity parameter
\begin{equation}
\xi = \frac{\gamma_0 \beta_0 a_0}{T_0} \,,
\end{equation}
and $K(k)$, the complete elliptic integral of the first kind. Moreover, the magnetic field and density profiles
can be expressed as
\begin{align}
  B_{0z}(y) &=  \frac{2 u T_0}{\gamma_0 \beta_0}
  \frac{\textrm{dn}\left[u \left(y -\lambda_0/4 \right),\kappa^2\right]}{\textrm{cn}\left[u \left(y -\lambda_0/4 \right),\kappa^2\right]} \frac{m  c^2}{e} \,, \label{eq:jacobi_B} \\
  n_{\alpha 0}(y) &= N \left\{\frac{\textrm{cn}\left[u \left(y -\lambda_0/4 \right),\kappa^2\right]}{1 \pm \textrm{sn}\left[u \left(y -\lambda_0/4 \right),\kappa^2\right]}\right\}^2 \,, \label{eq:jacobi_n}
\end{align}
where $\textrm{sn}(u,k)$, $\textrm{cn}(u,k)$ and $\textrm{dn}(u,k)$ are the Jacobi elliptic functions. The $\pm$ sign in Eq.~\eqref{eq:jacobi_n}
denotes the sign of the current of species $\alpha$. The parameters $u$ and $\kappa$ are defined as
\begin{align}\label{eq:params_harris}
  u &= 2 \frac{\gamma_0 \beta_0}{\sqrt{T_0}}  \sinh \left(\xi / 2 \right)   \frac{\omega_p}{c} \,, \\
  \kappa &= \coth \left( \xi/2 \right) \,.
\end{align}

A given species is in a strongly nonlinear (pinched) regime if $\xi_\alpha \equiv \gamma_\alpha \beta_\alpha a_0/T_\alpha \gg 1$ (\emph{i.e.},
$\kappa \to 1^+$), in which case its peak density greatly exceeds $N_\alpha$. If this condition is fulfilled by all species, the system then
consists of a periodic chain of well-separated neutral current sheets, which tend to the Harris solution \cite{Harris_1962, Zelenyi_1979, Balikhin_2008},
widely used in magnetic reconnection studies. This result readily follows from taking the $\kappa \to 1^+$ limit of the Jacobian elliptic functions \cite{Byrd_1954} involved in Eqs.~\eqref{eq:jacobi_B} and~\eqref{eq:jacobi_n} around the density maximum. Applying the latter limit to~\eqref{eq:jacobi_n} leads to 
\begin{align}
  n_0(y) &\simeq \frac{N}{4} \exp(u \lambda_0/2) \cosh^{-2}(u y) \\
& \simeq N e^\xi \cosh^{-2}(u y) \,,
\end{align}
which corresponds to a Harris-type relativistic current sheet \cite{Zelenyi_1979, Balikhin_2008} with maximum density
\begin{equation}
  \max_y n_0 = N e^\xi \,,
\end{equation}
and characteristic width
\begin{align} \label{eq:Harris_width}
  l &= u^{-1} \\
  &\simeq \frac{\sqrt{T_0} e^{-\xi/2}}{\gamma_0 \beta_0} \frac{c}{\omega_p}
  =  \frac{\sqrt{T_0}}{\gamma_0 \beta_0} \frac{c}{\Omega_p} \,.
\end{align}
In the same limit, Eq.~\eqref{eq:jacobi_B} becomes (for $l \ll y \ll \lambda_0$)
\begin{align}
  \frac{B_{0z}^2(y)}{8 \pi} &\simeq 2 T_0 mc^2 N e^\xi \\
  & \simeq 2 T_0 mc^2 \max_y n_0 \,,
  \label{eq:press_bal}
\end{align} 
which expresses the expected balance between magnetic and thermal pressures in a current filament. 

\subsection{\label{sec:level2_pert}Perturbative system}

We now linearize the fluid-Maxwell equations \eqref{eq:continuity}-\eqref{eq:Gauss} around the initial state described by
Eqs.~\eqref{eq:momentum_stat}-\eqref{eq:Gauss_stat}. Since the steady state is invariant along $x$, the first-order quantities are
taken in the form $\delta b(y)e^{ik_x x - i\omega t}$, where $k_x$ is the longitudinal wavenumber, $\omega = \omega_r + i\Gamma$
is the complex frequency and $\delta b(y)$ is the eigenfunction. The fluid equations are closed assuming an adiabatic equation of state
\begin{equation}
  p_\alpha \propto n_\alpha^{\Gamma_\mathrm{ad,\alpha}} \,.
\end{equation}
In the following, we will take $\Gamma_\mathrm{ad, \alpha}=4/3$ in a relativistically hot beam ($T_\alpha \gtrsim 1$) and  $\Gamma_\mathrm{ad,\alpha}=5/3$
otherwise.

The linearized system, detailed in Appendix~\ref{app:fluid_equations}, consists of 10 independent coupled homogenous differential equations
with $y$-periodic coefficients that depend on the equilibrium quantities. This system can be written in the compact form
\begin{equation}\label{eq:syst}
  \partial_y \bm{x}(y) = \bm{\Xi}(y,\omega,k_x) \bm{x}(y) \,,
\end{equation}
where $\bm{x}(y) = (\delta e_x,\delta b_z,\delta p_1,...\delta p_4,\delta v_{y1},...,\delta v_{y4})$ is a 10-component vector containing
the first-order variables, and $\bm{\Xi}(y,\omega,k_x)$ is a $10\times10$ $y$-periodic matrix. This equation
can be solved using the Floquet theory \cite{Grimshaw_1993} as explained in the next section.  

\section{\label{sec:level1_Floquet}Floquet analysis}

\subsection{\label{general_method}General method}

We now recall the basics of the Floquet theory \cite{Grimshaw_1993} used to solve Eq.~\eqref{eq:syst}. For the sake of clarity,
the vectors and matrices are written in lowercase and uppercase, respectively. Equation~\eqref{eq:syst} has the general form
\begin{equation}\label{eq:floquet}
  \partial_y \bm{x}(y) = \bm{\Xi}(y) \bm{x}(y) \,, 
\end{equation}	
where $\bm{x}$ is a $n$-dimensional vector and $\bm{\Xi}(y)$ is a $n\times n$ periodic matrix, \emph{i.e.},
\begin{equation}\label{eq:periodic}
  \bm{\Xi}(y) = \bm{\Xi}(y+\lambda),\ \forall y \in \mathbb{R}.
\end{equation}
Here, $\lambda$ is the minimal periodicity of the system satisfying the relation (\ref{eq:periodic}). We now introduce the fundamental matrix
\begin{equation}\label{eq:M}
  \bm{M}(y) = \Big[ \left[\bm{x}_1(y)\right]...\left[\bm{x}_n(y)\right] \Big] \,,
\end{equation}
where $\bm{x}_j(y),...,\bm{x}_n(y)$ are $n$ linearly independent solutions to Eq.~\eqref{eq:floquet}. $\bm{M}(y)$ satisfies the 
differential equation
\begin{equation}\label{eq:M_syst}
  \partial_y \bm{M}(y) = \bm{\Xi}(y) \bm{M}(y) \,.
\end{equation}
According to the Floquet theory, $\bm{M}$ has the form
\begin{equation}\label{eq:solution}
  \bm{M}(y) = \bm{P}(y) \bm{\Sigma}(y) \,,
\end{equation}
where
\begin{equation}\label{eq:rho}
  \bm{\Sigma}(y) = \textrm{diag}\Big[ e^{\mu_1 y},...,e^{\mu_n y} \Big ] \,,
\end{equation}	
with $\mu_j,...,\mu_n$ the so-called characteristic Floquet exponents. The matrix $\bm{P}(y)$ is composed of $n$ vector-valued periodic functions:
\begin{equation}\label{eq:P}
\bm{P}(y) = \Big[ \left[\bm{p}_1(y)\right]... \left[\bm{p}_n(y)\right] \Big]
\end{equation}
such that
\begin{equation}\label{eq:P_per}
\bm{p}_j(y) = \bm{p}_j(y+\lambda),\ \forall y \in \mathbb{R}.
\end{equation}
The real parts of the Floquet exponents are well defined, yet their imaginary parts are not unique since a phase shift of $2 \pi/\lambda_0$ can be applied indifferently to the Floquet exponents or $\bm{P}(y)$.

The diagonal elements of $\bm{\Sigma}(\lambda) = \textrm{diag}[\sigma_1,...,\sigma_n]$ are called the characteristic multipliers,
which can be obtained as the eigenvalues of the non-singular constant matrix $\bm{B}$ defined by
\begin{equation}\label{eq:B}
  \bm{B} = \bm{M}^{-1}(y) \bm{M}(y+\lambda) \,.
\end{equation}
Introducing $\bm{b}_1,...,\bm{b}_n$ the eigenvectors of $\bm{B}$, and defining the matrix
\begin{equation}\label{eq:S}
  \bm{S} = \Big[\left[\bm{b}_1\right]...\left[\bm{b}_n\right] \Big] \,,
\end{equation}
we have
\begin{equation}\label{eq:rho_lambda}
  \bm{\Sigma}(\lambda) = \bm{S}^{-1} \bm{B} \bm{S} \,,
\end{equation}
so that
\begin{equation}\label{eq:P_S}
  \bm{P}(y) = \bm{\Sigma}(y)\ \bm{S}\ \bm{\Sigma}^{-1}(y) \, .
\end{equation}
An important result is that the eigenvectors and eigenvalues of $\bm{B}$ do not depend on the choice of initial conditions. For a given set of 
$\bm{x}_j(y),...,\bm{x}_n(y)$, numerically computed over a period, we evaluate $\bm{B}=\bm{M}^{-1}(0)\bm{M}(\lambda)$ and solve for its eigenvalues
and eigenmodes. In practice, for a given value of $k_x \in \mathbb{R}$, we scan a region of the complex $\omega$-space, and retain only the temporally
unstable eigenmodes, \emph{i.e.}, those with purely real transverse wavenumbers, $k_y \equiv -i\mu_i \in \mathbb{R}$. Inverting the results readily gives the dispersion relation $\omega(k_x,k_y)$. This technique, previously
used in Ref.~\cite{Romeiras_1986}, has the advantage of not requiring the infinite Hill's determinant to be derived, as is commonly done when applying
the Floquet theory to systems periodic in space time (see, e.g., \cite{Quesnel_1997}).

In order to gauge the accuracy of our results, we check that the exact relation
\begin{equation}\label{eq:detB}
  \textrm{det}\ \bm{B} = \textrm{exp}\left\{\int_{0}^{\lambda} \Tr \left[ \bm{\Xi}(y) \right] dy\right\} \,.
\end{equation}
is well satisfied. Also, due to the possible presence of spatially fast-growing modes ($\Im k_y \ne 0$), we use initial conditions such that
$\bm{x}_j^i(\lambda/2)=\delta_j^i$  ($i,j=1,...,n$).

\subsection{\label{subsec:level1_homo}Instability of two homogeneous symmetric counterstreaming pair beams}

We first address the problem of a homogeneous plasma system (or, equivalently, vanishing fields), in which case we expect to recover the
standard linear dispersion relation \cite{Bret_2010a}. Specifically, we consider two symmetric counterstreaming pair beams with $T_0=1$ and $\gamma_0=10$.
Owing to the symmetry of the system, the complex frequency is purely imaginary ($\omega = i\Gamma$), and hence the search for the $\omega$ values yielding
$k_y \in \mathbb{R}$ is restricted to a 1D parameter space (at fixed $k_x$). The results of the Floquet solver are compared with those obtained from the
homogeneous linear dispersion relation. This dispersion relation, derived from linearizing Eqs.~\eqref{eq:continuity}-\eqref{eq:Gauss} in the homogeneous limit,
takes on the form of a 10th-order polynomial. This relation is too lengthy to be written here, but a more compact expression of it can be obtained using the covariant formalism developed in Ref. \cite{Achterberg_2007_I}. For each point in the $(k_x,k_y)$ Fourier space, we compute the roots $\omega(k_x,k_y)$ of this polynomial and
retain the solution with maximum (and positive) $\Im \omega$. For the Floquet solver, we sample the parameter space $(\Gamma, k_x) \in \mathbb{R}^+\times \mathbb{R}^+$,
and look for eigenmodes with $k_y \in \mathbb{R}$.

\begin{figure}[t!]
\begin{tabular}{c}
  \includegraphics[width=0.45\textwidth]{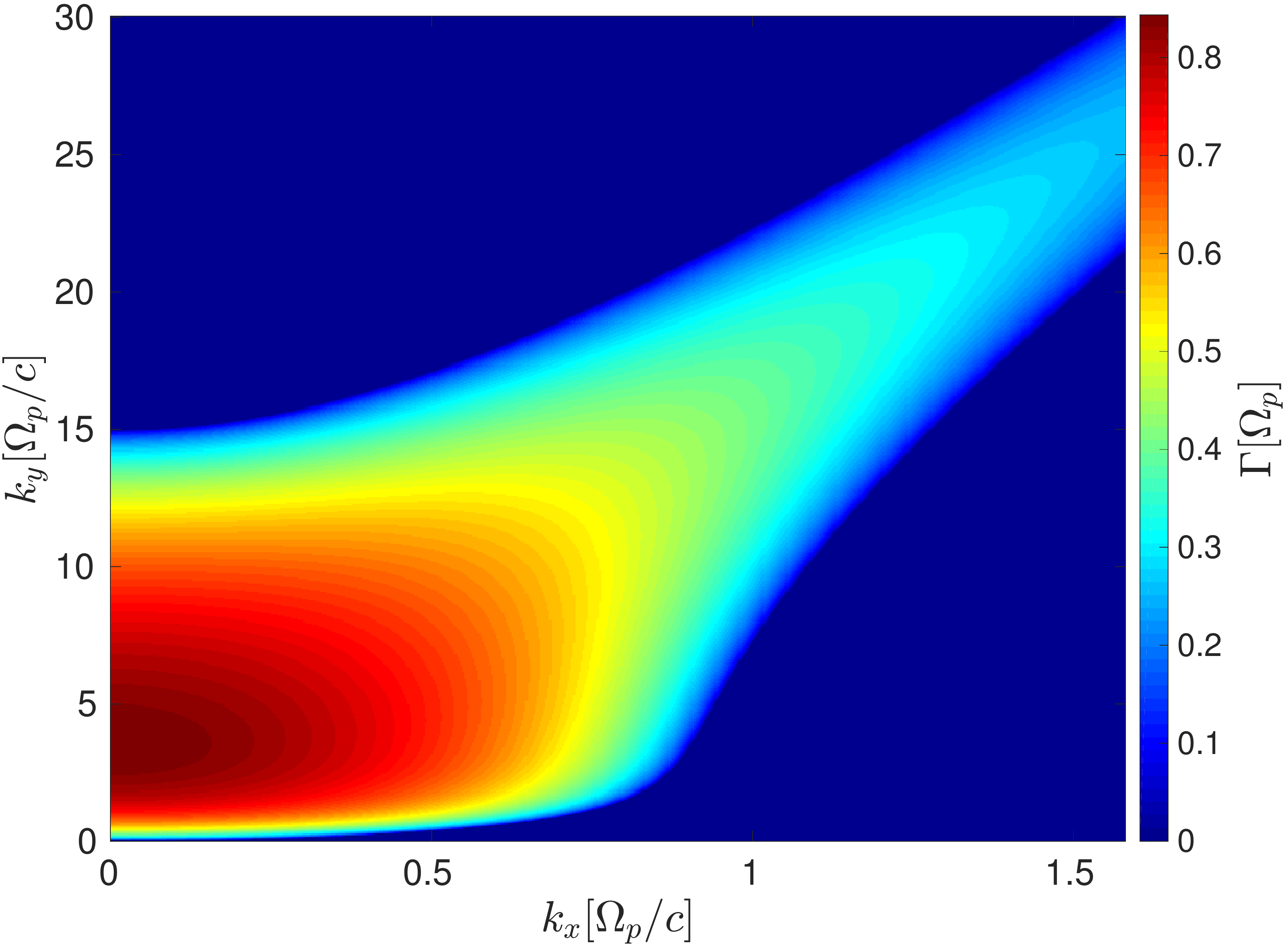} \\
  \includegraphics[width=0.45\textwidth]{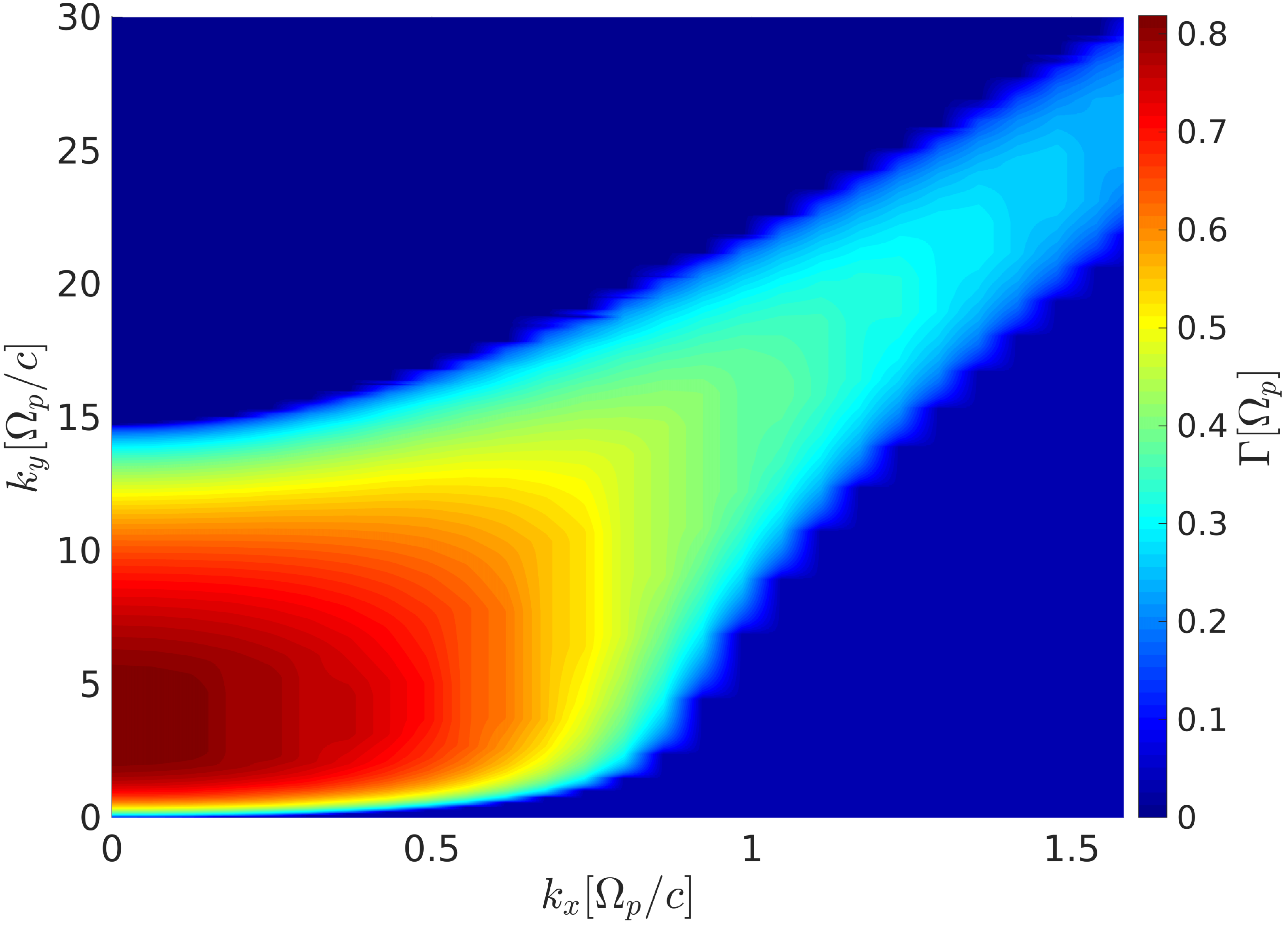}
\end{tabular}
  \caption{Instability growth rates ($\Gamma$) in the $(k_x,k_y)$ plane for two symmetric counterstreaming pair beams with $T_0 = 1$ and $\gamma_0 = 10$,
  as obtained from the linear dispersion relation (top) and the Floquet method (bottom).\label{fgr:lin_GR}}
\end{figure}

Figure~\ref{fgr:lin_GR} confirms that the two methods give identical growth-rate maps. In the Floquet case (bottom panel), the cruder resolution in $k_y$
follows from the chosen resolution in $\Gamma$. As expected \cite{Bret_2010a}, the system gives rise to a broad unstable spectrum, characterized by
dominant, purely tranverse CFI modes (with $k_x=0$) and a continuum of sub-dominant oblique modes extending up to infinite $k$.
A more physical kinetic description would evidently lead to a bounded unstable spectrum due to Landau damping of high-$k$ modes \cite{Bret_2010a}.

\section{\label{sec:level1_weak}1D instability of weakly nonlinear symmetric filaments}

The purely transverse CFI is the dominant instability arising in a symmetric relativistic two-beam plasma in the homogeneous limit \cite{Bret_2010a}.
As will be shown in \ref{sec:level1_strong}, the dominant unstable mode in a periodic system of weakly pinched filaments remains purely
transverse ($k_x=0$), and can be viewed as a filament merging instability (FMI). In this section, we examine the properties of this instability in a
purely 1D geometry. 

\subsection{\label{sec:level2_weak_1}Floquet analysis}

\begin{figure}[t!]
  \centering
  \includegraphics[width=0.9\hsize]{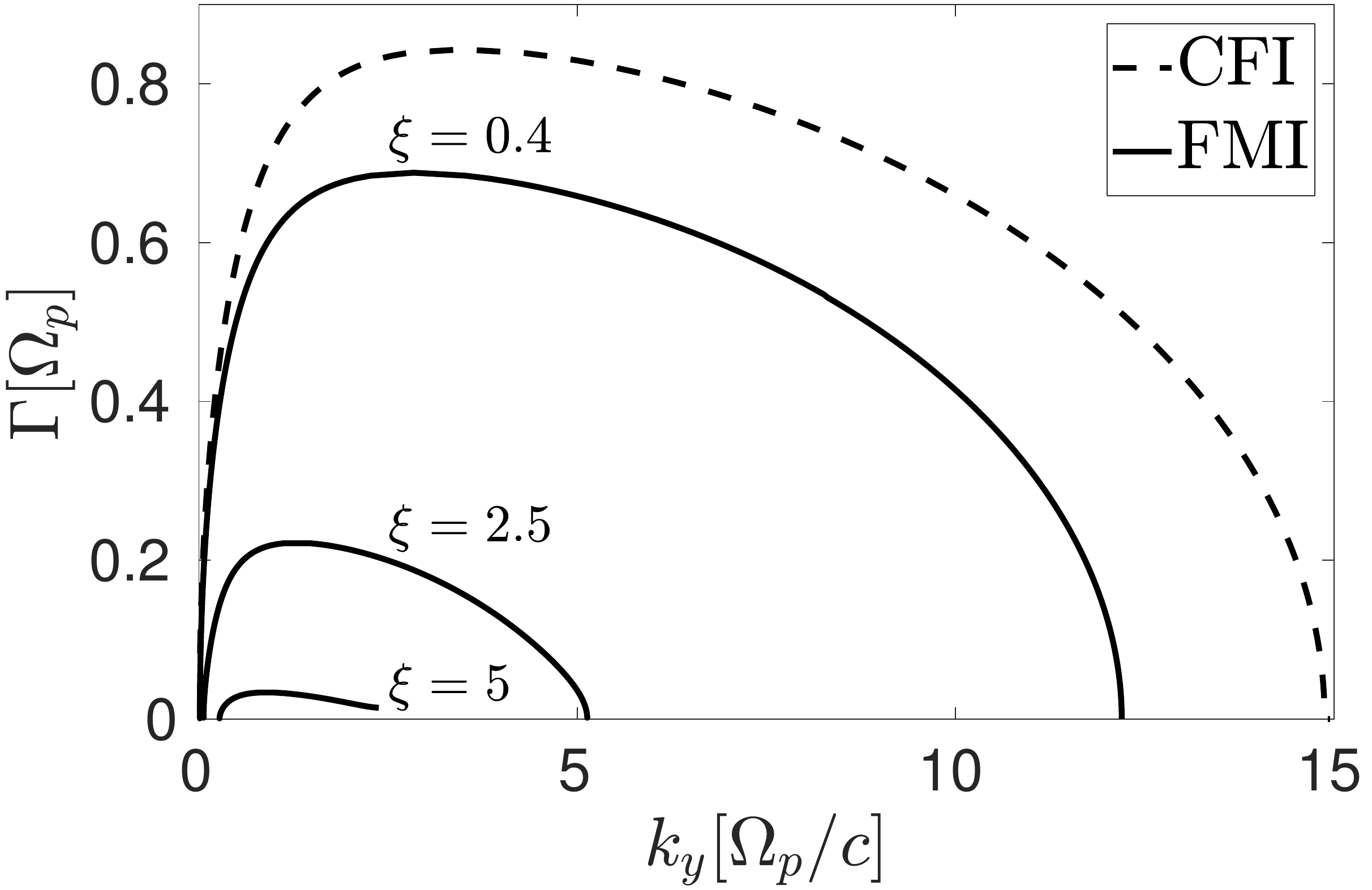}
  \caption{Growth rate ($\Gamma$, solid curve) of the FMI as a function of the characteristic Floquet exponent ($k_y$).
  The unperturbed periodic system consists of two symmetric counterstreaming pair beams with $T_0=1$, $\gamma_0=10$.
   Three different field strengths, $a_0=0.04, 0.25, 0.5$ (\emph{i.e.}, $\xi = 0.4, 2.5, 5$), are considered, which are associated with fundamental wavenumbers $k_0=16.5, 8.03, 4.88\,\Omega_p/c$ (\emph{i.e.}, $\lambda_0=0.38,  0.78, 1.28\,c/\Omega_p$). For comparison,
  the dashed curve plots the growth rate of the CFI in the homogeneous limit  with the same ($T_0,\gamma_0$) parameters.}
  \label{fgr:disp_rel_sym}
\end{figure}

We study the stability of a stationary symmetric two-beam system characterized by $T_0=1$, $\gamma_0=10$ and $a_0=0.04$ ($\xi = 0.4$),
associated with a fundamental wavelength $\lambda_0 \simeq 0.38\,c/\Omega_p$. We solve the perturbative eigenvalue problem using the method
previously explained with $k_x=0$ and $\omega = i\Gamma$ purely imaginary. For each sampled value of $\Gamma$, we find two solutions
with real positive $k_y$, and their symmetric negative counterparts (corresponding to the same physical modes). Figure~\ref{fgr:disp_rel_sym}
plots the growth rate $\Gamma$ as a function of the characteristic Floquet exponent $k_y \in [0,2\pi/\lambda_0]$. Since $\Gamma (k_y)$ is even,
it should remain unchanged under the operation $k_y \rightarrow k_y + 2n\pi/\lambda_0$ ($n\in \mathbb{N}$). For clarity, we have chosen to unfold
$\Gamma(k_y)$ over the range $[0,2\pi/\lambda_0]$, but one could have alternatively chosen the range $[0,\pi/\lambda_0]$, yielding a
double-valued curve. Unstable modes are found up to $k_y \simeq 12.2\,\Omega_p/c$, to be compared with the fundamental wavenumber
$k_0 \equiv 2\pi/\lambda_0 \simeq 16.5\,\Omega_p/c$. The maximum growth rate, $\Gamma_\mathrm{max} = 0.69\,\Omega_p$, occurs at
$k_{y, \mathrm{max}} \simeq 2.85\,\Omega_p/c$. We also plot the CFI growth rates associated with the same values of $T_0$ and $\gamma_0$.
The two curves present similar shapes, but the plasma inhomogeneity tends to reduce both the maximum growth rate and wavenumber range of the instability.

\begin{figure}[t!]
  \centering
  \includegraphics[width=0.9\hsize]{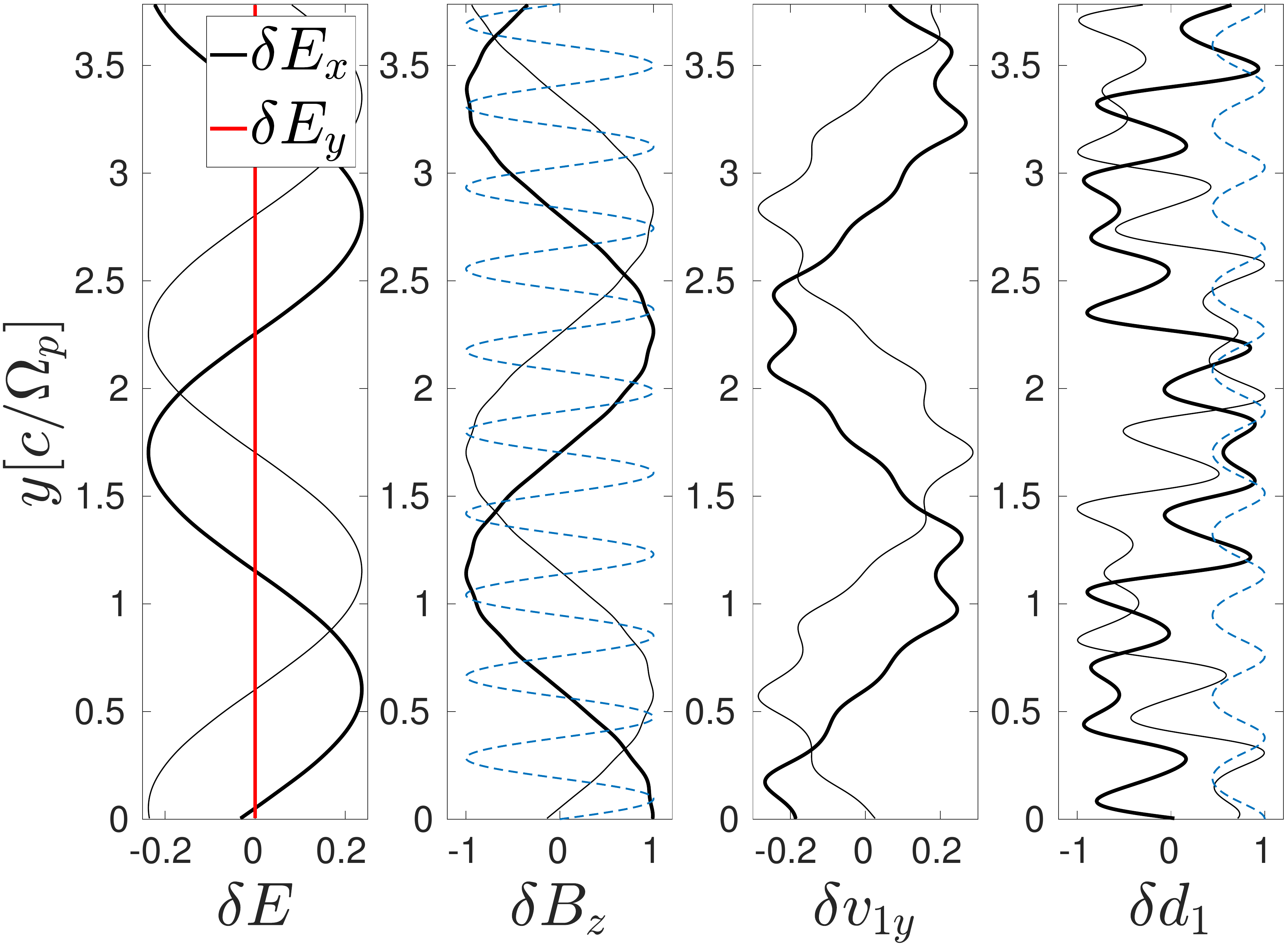}
  \caption{Spatial structure of the dominant FMI eigenmode for the parameters of Fig.~\ref{fgr:disp_rel_sym}. The thick (resp. thin) solid lines
  plot the real (resp. imaginary) parts of the eigenfunctions. The perturbed quantities are, from left to right, the inductive electric field
  ($\delta E_x$, black), the electrostatic electric field ($\delta E_y$, red), the magnetic field ($\delta B_z$), the transverse velocity
  ($\delta v_{1y}$), and the lab-frame density of plasma species 1 ($\delta d_1$). $\delta E_x$, $\delta E_y$ and $\delta B_z$ are normalized to
  $\max_y \vert \delta B_z \vert$, while $\delta v_{1y}$ and $\delta d_1$ are normalized to $\max_y \vert \delta d_1 \vert$. The blue dashed lines plot the
  corresponding unperturbed quantities normalized to their maximum value.}
  \label{fgr:prof_o1}
\end{figure}

The spatial profiles of the components $(\delta E_x, \delta E_y, \delta B_z, \delta v_{1y}, \delta d_1)$ of the fastest-growing FMI eigenmode are
displayed in Fig.~\ref{fgr:prof_o1}. Due to the symmetric configuration, only the perturbation of the positron fluid with positive velocity are shown. Both the real and imaginary parts of the eigenfunctions are plotted, showing that they only differ by a
constant phase of $\sim \pi/2$. The electromagnetic field perturbations $(\delta E_x, \delta B_z)$ are well described by a single
harmonic function with wavenumber $k_{y,\mathrm{max}}$, so that $\vert \delta E_x /\delta B_z \vert \simeq \Gamma_\mathrm{max}/k_{y, \rm max} \simeq 0.26$.
By contrast, $\delta v_{1y}$, $\delta d_1$ show a richer harmonic content, with modulations at the scale of $\lambda_0$. Yet the charge density modulations
cancel out to yield a zero electrostatic field $\delta E_y$.

\begin{figure}[t!]
  \centering
  \includegraphics[width=0.8\hsize]{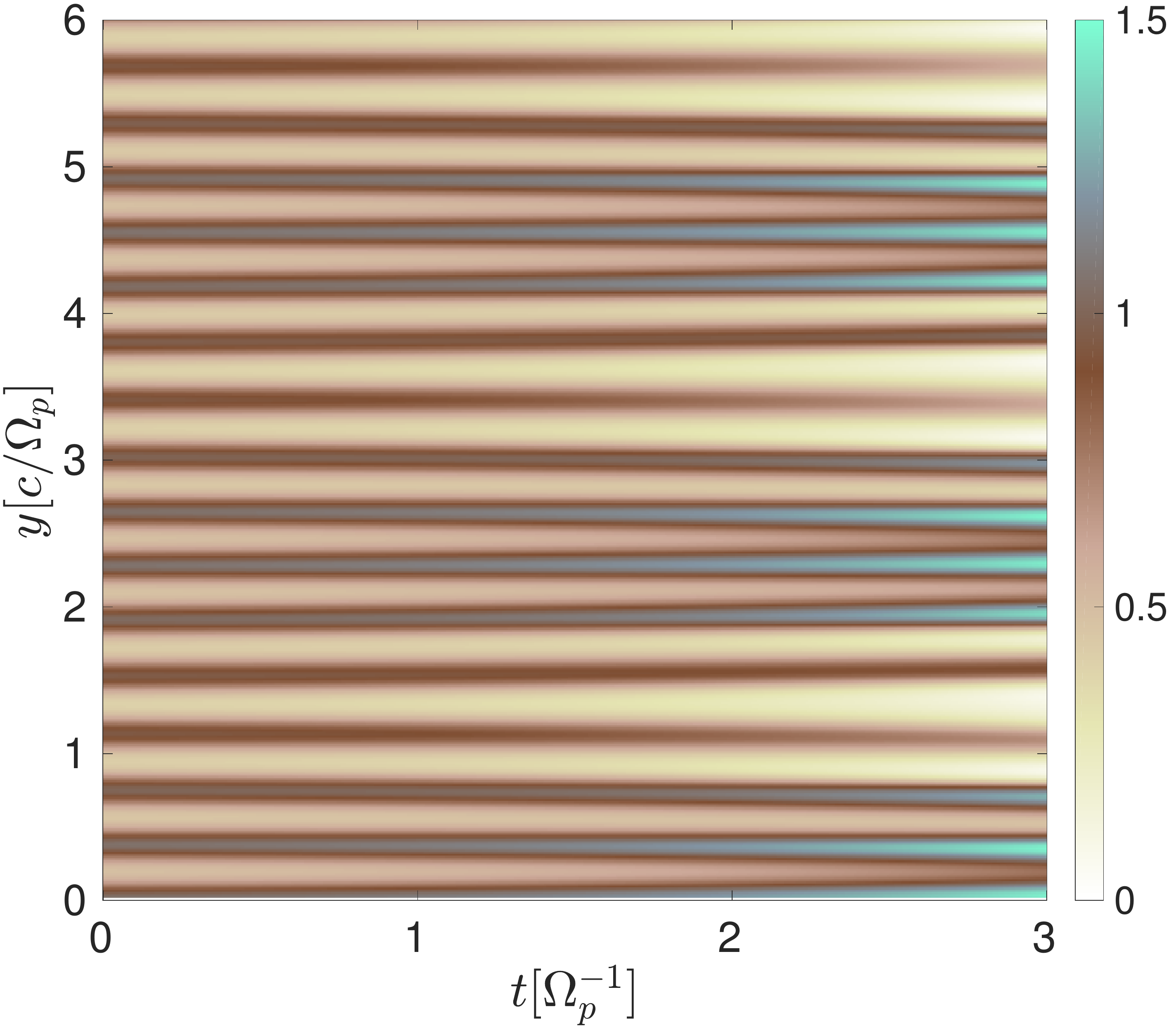}
  \caption{Space-time evolution of the density component, $d_1 = d_{10}(y)+\delta d_1(y)e^{\Gamma t}$ (normalized to $\mathrm{max}_y d_{10}$),
  of the fastest-growing FMI eigenmode. The parameters are those of Fig.~\ref{fgr:disp_rel_sym}. The evolution is stopped slightly
  before $d_1$ becomes negative.}
  \label{fgr:fil_weak}
\end{figure}

The space-time evolution of the dominant FMI mode is depicted in Fig.~\ref{fgr:fil_weak} by plotting the total apparent density of species 1, 
$d_1(y,t) = d_{10}(y) + \epsilon \delta d_1(y) e^{\Gamma t}$ (where $\epsilon \ll 1$). For the sake of being illustrative, $d_1(y,t)$ is evolved
up to the point where it vanishes, \emph{i.e}, strictly speaking beyond the linear regime. One clearly sees that the instability has the effect
of creating bunches of three or four filaments, whose density rises as they coalesce. The filaments surrounding the merging ones also experience
magnetic attraction, albeit weaker. In the nonlinear saturated stage, the system will then be reorganized into a quasiperiodic filamentary pattern with
a characteristic wavelength about 6 times longer than in its unperturbed state.

\subsection{\label{sec:level2_weak_2}1D PIC simulation}

To support the results of the Floquet analysis, we have performed a 1D3V PIC simulation (1D in space and 3D in momentum) using the code \textsc{calder}
\cite{Lefebvre_2003}. In accordance with the theoretical calculation, the plasma comprises two counterstreaming pair beams, made up of a total of
four species.  Each species $\alpha$ is initialized with a drifting J\"uttner-Synge momentum distribution:
\begin{align}\label{juttner}
  f_{\alpha 0}(y,\bm{p}) &= N_{\alpha} \exp\left\{-\frac{\gamma_{\alpha 0}}{T_{\alpha 0}} \left[\gamma(\bm{p})+\frac{q_\alpha \phi_0(y)}{m c^2} \right. \right. \nonumber \\
  &\left. \left. -\frac{\beta_\alpha}{m c} \left( p_x +\frac{q_\alpha}{c} \beta_{\alpha 0} A_{0x}(y) \right) \right] \right\} \,,
\end{align}
where the equilibrium potentials $(\phi_0, A_{0x})$ are numerical solutions to Eqs.~\eqref{eq:A_stat} and \eqref{eq:phi_stat}. In the present
symmetric case, we have $\phi_0(y)=0$, and hence the initial electromagnetic fields are $B_{0z}(y)=-\partial_y A_{0x}$ and $E_{0x}(y)=E_{0y}(y)=0$.
The physical parameters are those considered in the previous section: $T_0=1$, $\gamma_0=10$ and $a_0=0.04$, giving $\lambda_0 = 0.38c/\Omega_p$.
The simulation domain is $50\,\lambda_0$ long, allowing the unstable modes to be accurately resolved. The mesh size is
$\Delta y = \lambda_0/100 = 0.0038c/\Omega_p$ and the time step is
$\Delta t = 0.99\, \Delta x/c$. Each cell initially contains 1000 macroparticles per species. Periodic boundary conditions are used for both fields and particles. The instability is seeded by thermal noise alone.

\begin{figure}[t!]
  \centering
  \includegraphics[width=0.8\hsize]{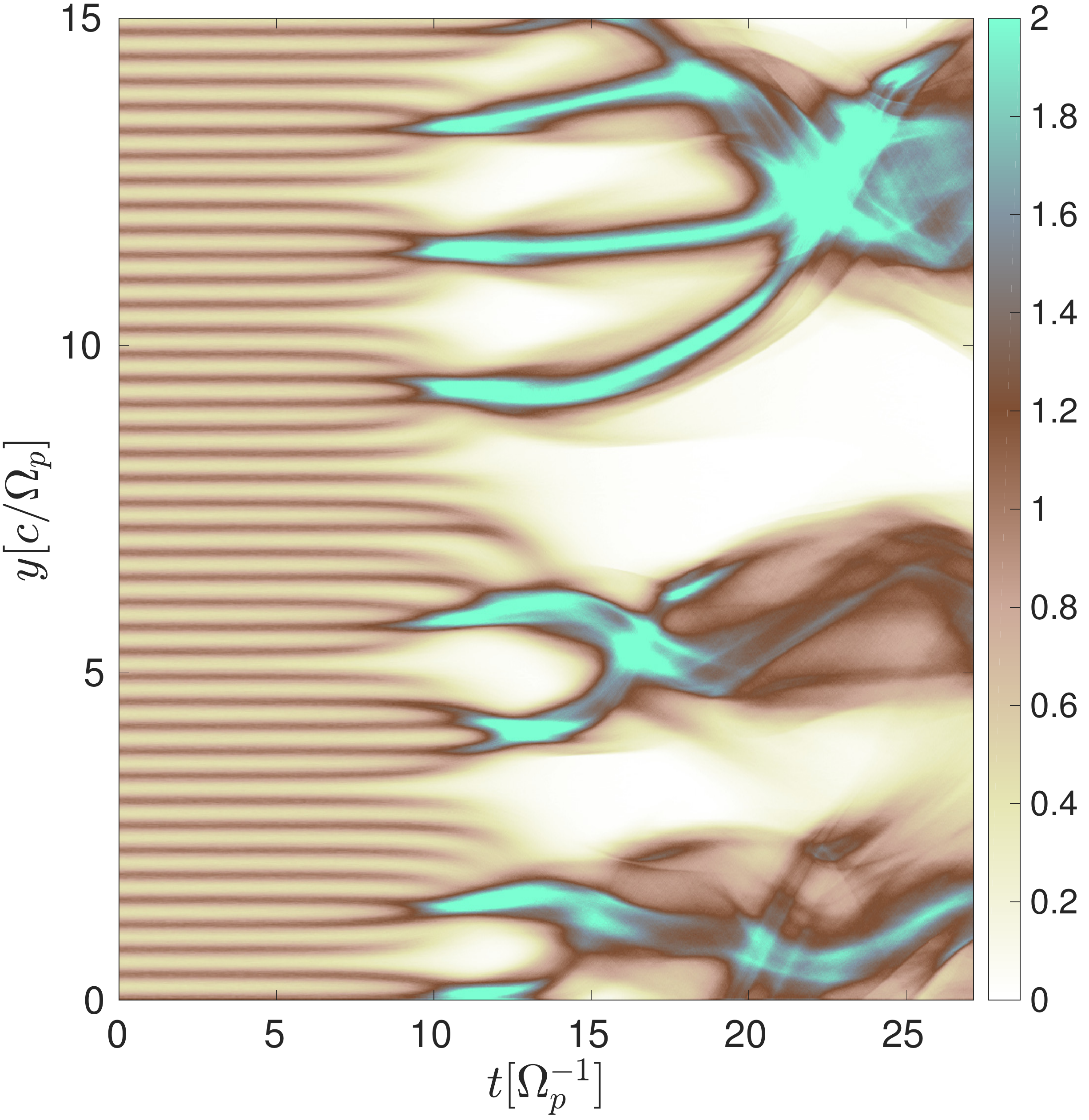}
  \caption{1D PIC simulation with the parameters $T_0=1$, $\gamma_0=10$ and $a_0=0.04$ (as in Fig.~\ref{fgr:disp_rel_sym}):
  space-time evolution of the density of plasma species 1 (normalized to its initial peak value).}
  \label{fgr:fil_merge}
\end{figure}

Figure~\ref{fgr:fil_merge} displays the space-time evolution of the density of species 1 (positrons moving along $x>0$). A first instability
stage is observed to last until $t \lesssim 10\,\Omega_p^{-1}$: it is characterized by mergers between clusters of 3-4 neighboring filaments, consistent
with the theoretical prediction of Fig.~\ref{fgr:fil_weak}. The larger and denser filaments resulting from this primary instability undergo successive
merging stages, leading to increasingly wide and spaced filaments.

\begin{figure}[t]
  \centering
  \includegraphics[width=0.8\hsize]{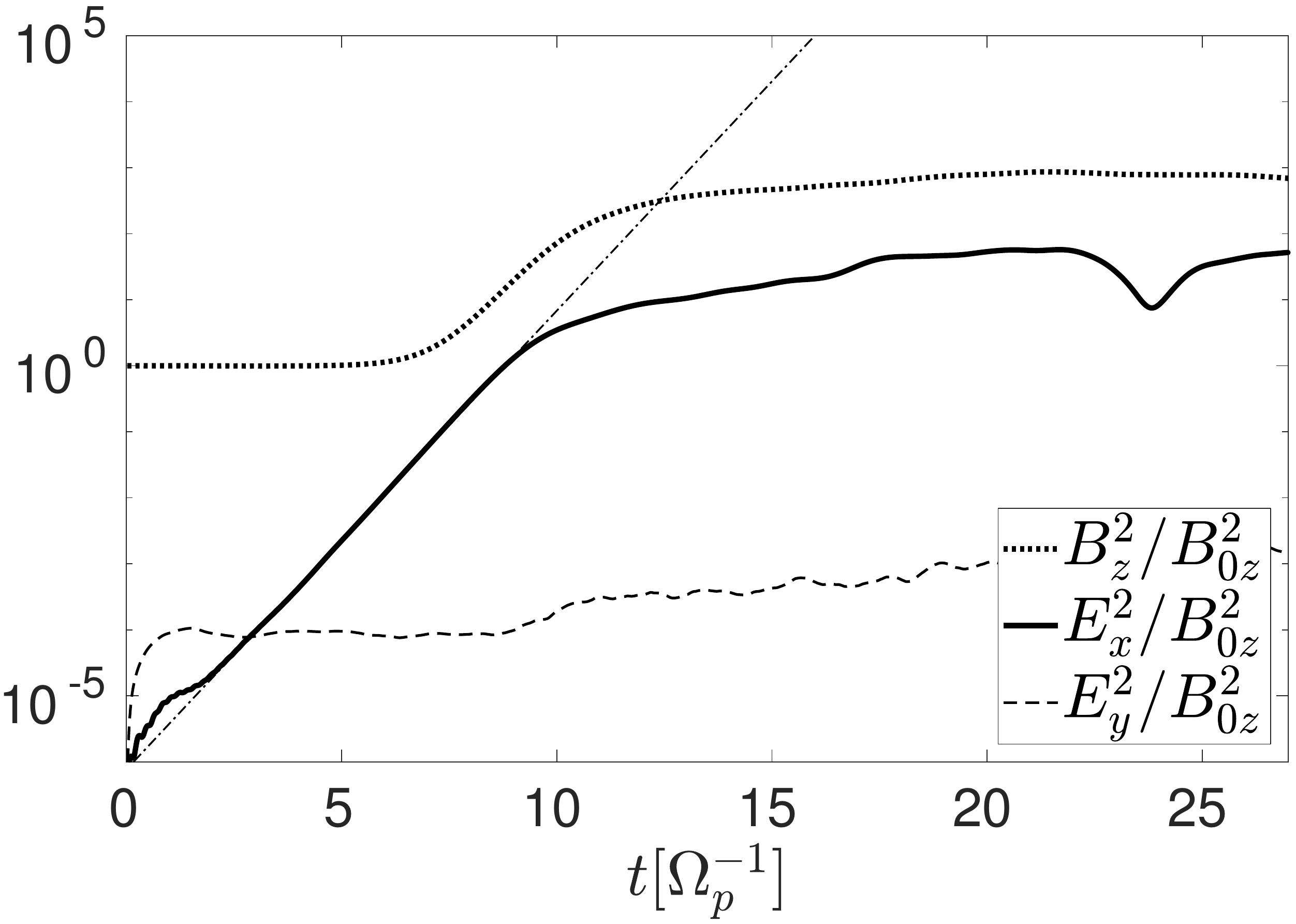}
  \caption{Time evolution of the electromagnetic energies (spatially integrated and normalized to the initial magnetic energy). 
   The initial simulation parameters are those of Fig.~\ref{fgr:fil_merge}. The magnetic ($B_z$) energy is plotted as a dotted line, the inductive
   ($E_x$) energy as a thick solid line and the electrostatic ($E_y$) energy as a dashed line. The thin dotted-dashed line represents the best-fitting
   exponential curve over $5 \le t \le 10\,\Omega_p^{-1}$, with a growth rate $\Gamma_\mathrm{PIC} \simeq 0.8\,\Omega_p$.}
  \label{fgr:fields_weak}
\end{figure}

\begin{figure}[t]
  \centering
  \includegraphics[width=0.8\hsize]{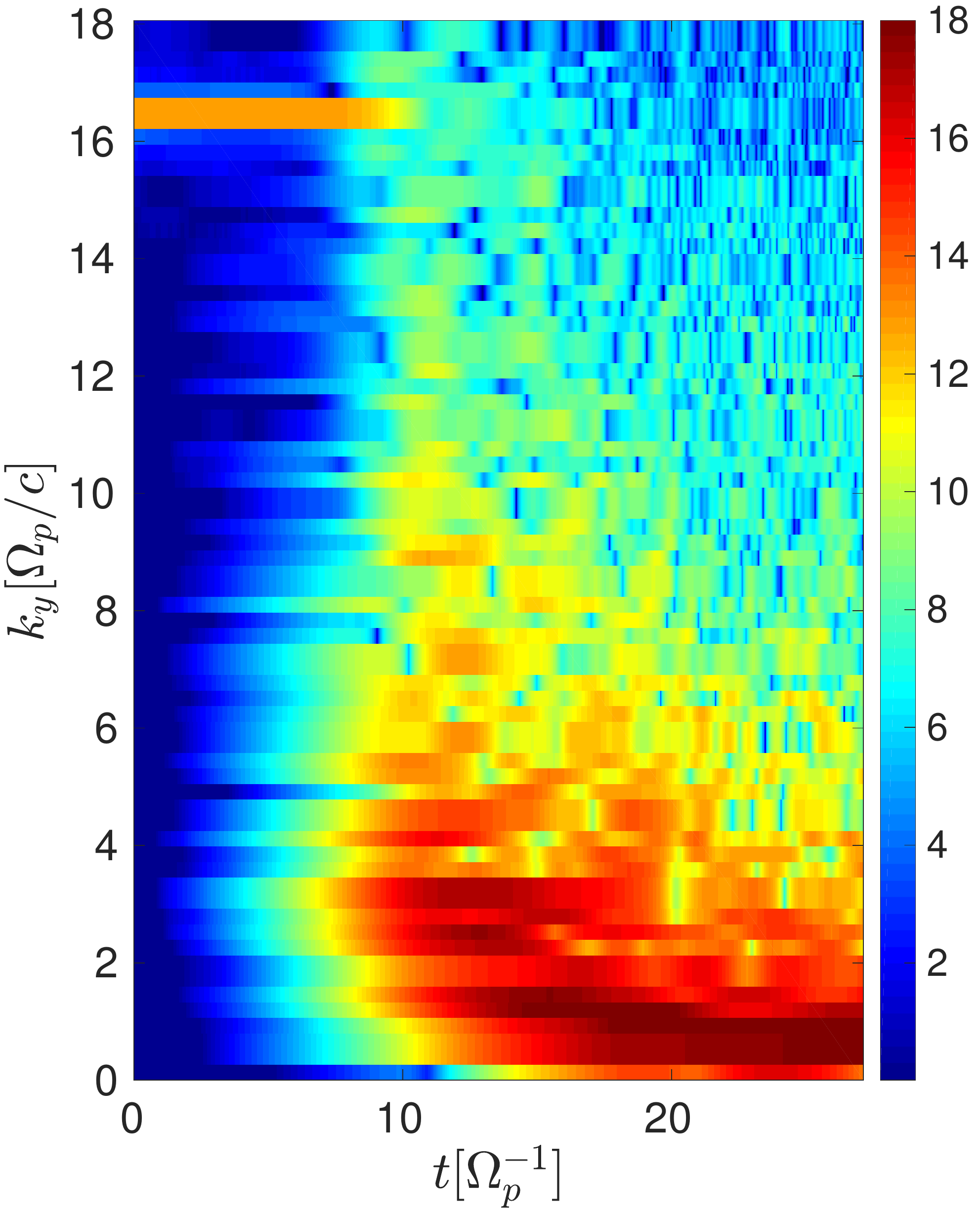}
  \caption{Time evolution of the Fourier transformed magnetic field, $\vert B_z(k_y,t)\vert ^2$ (in $\log_{10}$ scale). The initial simulation
  parameters are those of Fig.~\ref{fgr:fil_merge}. The bright spectral line at $k_y\simeq 16.5\,\Omega_p/c$ corresponds to the fundamental
  wavenumber of the equilibrium state.
  \label{fgr:fft}}
\end{figure}

The linear properties of the early-time FMI can be further quantified from the time evolution of the total electromagnetic energies (Fig.~\ref{fgr:fields_weak})
and of the Fourier transformed magnetic field $\vert B_z (k_y,t)\vert^2$ (Fig.~\ref{fgr:fft}). In Fig.~\ref{fgr:fields_weak}, the instability is seen to emerge
from the electric thermal noise at $t \simeq 2\,\Omega_{p}^{-1}$, at which time the inductive electric field ($E_x$) energy starts to rise exponentially. Since
the magnetic field energy is initialized at a finite level (contrary to the initially vanishing electric field), the effect of the instability on it is only
discernible after $t\simeq 7\,\Omega_p^{-1}$. Both $E_x$ and $B_z$ energies then grow as $\propto e^{2\Gamma_\mathrm{PIC}t}$ where $\Gamma_\mathrm{PIC} \simeq 0.80\,\Omega_p$
is the best-fitting growth rate over $4 \le t \le 8\,\Omega_p^{-1}$. This value is comparable with the theoretical prediction $\Gamma_\mathrm{max} \simeq 0.69\,\Omega_p$,
the small difference being ascribed to kinetic effects. In particular, one may question the accuracy of the adiabatic closure relation: our choice of
$\Gamma_\mathrm{ad} = 4/3$ is somewhat dubious as the temperature $T_0=1$ is not highly relativistic, and, more importantly, as it implies three degrees of
freedom (whereas our PIC simulation is only 1D). We have checked that using $\Gamma_\mathrm{ad}=2$, the value expected in a relativistically hot 1D plasma,
gives a growth rate $\Gamma_\mathrm{max} \simeq 0.80\,\Omega_p$ (for $k_{y, \rm max} \simeq 2.3\,\Omega_p/c$), closer to the simulation results.
The eigenfunctions of the dominant mode are hardly modified by the choice of $\Gamma_{\rm ad, \alpha}$. They predict an order of magnitude difference in
the $B_z$ and $E_x$ energies, which is well reproduced in the simulation during the period $7 \lesssim t \lesssim 10\,\Omega_p^{-1}$ (Fig.~\ref{fgr:fields_weak}).
Also, the non-increasing $E_y$ energy is consistent with the theoretical prediction.

Figure~\ref{fgr:fft} shows that the FMI mainly develops in the wavenumber range $1\lesssim k_y \lesssim 6\,\Omega_p/c$, as expected
from theory (Fig.~\ref{fgr:disp_rel_sym}). Note that, in principle, the wavenumbers $k_y$ in this Fourier spectrum should not be directly
compared to the characteristic Floquet exponents (see Fig.~\ref{fgr:disp_rel_sym}) since the eigenfunctions are of the form
$\delta b(y) = \sum_{n\in \mathbb{N}} \delta b_n e^{i(k_y+nk_0) y}$, with $k_0$ being the fundamental wavenumber of the stationary state and $\delta b_0$ being not necessarily dominant. Yet for
the weakly nonlinear filaments considered, the magnetic-field eigenfunction is well approximated by a single harmonic term
$\delta B_z \sim \delta b_0 e^{ik_y y}$ (see Fig.~\ref{fgr:prof_o1}), so that the Fourier wavenumbers can be assimilated to good accuracy 
to the Floquet exponents. The growing modes appear to saturate at $t \simeq 10\,\Omega_p^{-1}$, causing depletion of the fundamental magnetic
mode (located at $k_0 = 16.5\,\Omega_p/c$). From this time onwards, the magnetic energy in the system is essentially contained
in the unstable spectral range ($k_y \lesssim 6\,\Omega_p/c$), leading to an exponential growth of the total magnetic energy (Fig.~\ref{fgr:fft}).
As time increases, the magnetic spectrum progressively shrinks to low $k_y$ due to successive filament mergers, which account for the modulations
of the $E_x$ energy in the saturated stage (Fig.~\ref{fgr:fields_weak}).

A rough estimate of the gain in magnetic field energy at saturation can be derived from Amp\`ere's equation expressed at the initial and saturation times: $B_0 \propto 2 \exp(-\xi) \sinh (\xi)  d_0 \lambda_0$ and $B_{\rm sat} \propto d_{\rm sat} \lambda_{\rm sat}$, with $d_{\rm sat}$ and $\lambda_{\rm sat}$ being, respectively, the apparent density of a given species and the typical wavelength at saturation. The factor $2 \exp(-\xi) \sinh (\xi)$ accounts for the significant overlap of the weakly inhomogeneous, counterstreaming flows in the initial state (such overlap is neglected at saturation) and the mean flow velocity is assumed to remain close to $c$. Making use of mass conservation, $d_{\rm sat} \lambda_{\rm sat} = d_0 \lambda_0 (k_0/k_{\rm max})$, one then predicts $(B_{\rm sat}/B_0)^2 \simeq \left( \exp (\xi)/2 \sinh(\xi) \right)^2 (k_0/k_{\rm max})^2$. Given $\xi = 0.4$ and the theoretical prediction $k_0/k_{\rm max} \simeq 16.5/2.85$, one obtains $(B_{\rm sat}/ B_0)^2 \simeq 120$, in fair agreement with the simulated value measured at $t \simeq 10\,\Omega_p^{-1}$ [Fig.~\eqref{fgr:fields_weak}].

\section{\label{sec:level1_strong}2D instability of strongly nonlinear symmetric filaments}

In this section, we push the stationary system far into the nonlinear regime, so that the equilibrium filaments get strongly pinched and the magnetic field
can no longer be approximated by a single harmonic. We will demonstrate that the increasing nonlinearity of the system causes its dominant eigenmode to
evolve from a purely tranverse FMI to a drift kink-type instability (DKI). To show this, we consider the same symmetric two-beam system as before
(with $T_0=1$ and $\gamma_0=10$), but increase the vector potential amplitude from $a_0=0.04$ to $a_0=0.5$ (\emph{i.e.}, the nonlinearity parameter
rises from $\xi = 0.4$ to $\xi=5$).

\subsection{\label{sec:level2_strong_1}Floquet analysis}

Figure~\ref{fgr:disp_rel_sym} shows that for $\xi=2.5$ the FMI growth rate exhibits a significantly reduced maximum value ($\Gamma_{\rm max} \simeq 0.22\,\Omega_p$ at $k_{y,\rm max} \simeq 1.29\,\Omega_p/c$) and upper cutoff `wavenumber' $k_y \simeq 5.1\,\Omega_p/c$. Also, it presents a nonzero lower cutoff $k_y \simeq 0.06\,\Omega_p/c$ (hardly visible in Fig.~\eqref{fgr:disp_rel_sym}). When the nonlinearity is further strengthened ($\xi=5$), the lower cutoff increases to $k_y \simeq 0.27\,\Omega_p/c$ and the $\Gamma$ curve abruptly stops at $k_y = k_0/2 = 2.45\,\Omega_p/c$ (where $k_0 \simeq 4.88\,\Omega_p$ is the corresponding fundamental wavenumber). The latter discontinuity is a typical feature of unstable waves in periodic media (\emph{e.g.}, see \cite{Romeiras_1978,Romeiras_1986}). Actually, this gap becomes discernible for $\xi \gtrsim 3.5$, and results in a progressive quelling of the modes in the range $k_0/2 \le k_y \leq k_0$. Stabilization of the low-$k_y$ modes is also specific to the strongly inhomogeneous regime; this mechanism underlines a recently proposed scheme, which exploits density ripples to mitigate the CFI in the laser-plasma context \cite{Mishra_2014}. 

\begin{figure}[t!]
	\centering
	\includegraphics[width=0.8\hsize]{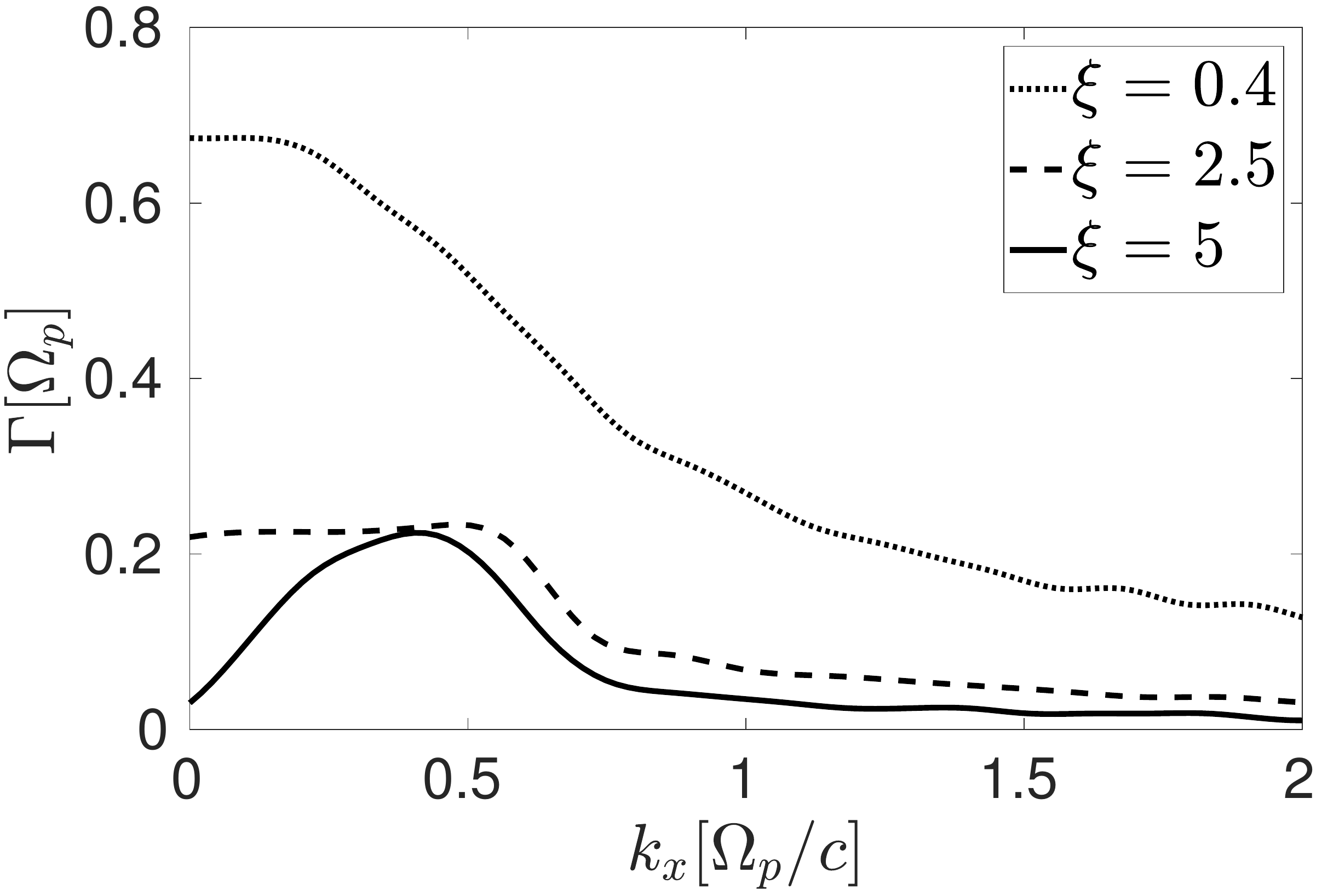}
	\caption{Growth rate $\Gamma$, maximized over $k_y$, as a function of the longitudinal wavenumber $k_x$ and the nonlinearity level $\xi$. 
		The symmetric two-beam system is characterized by $T_0 = 1$ and $\gamma_0 = 10$.}
	\label{fgr:disp_non_lin_tot}
\end{figure}

\begin{figure}[t!]
	\centering
	\includegraphics[width=0.8\hsize]{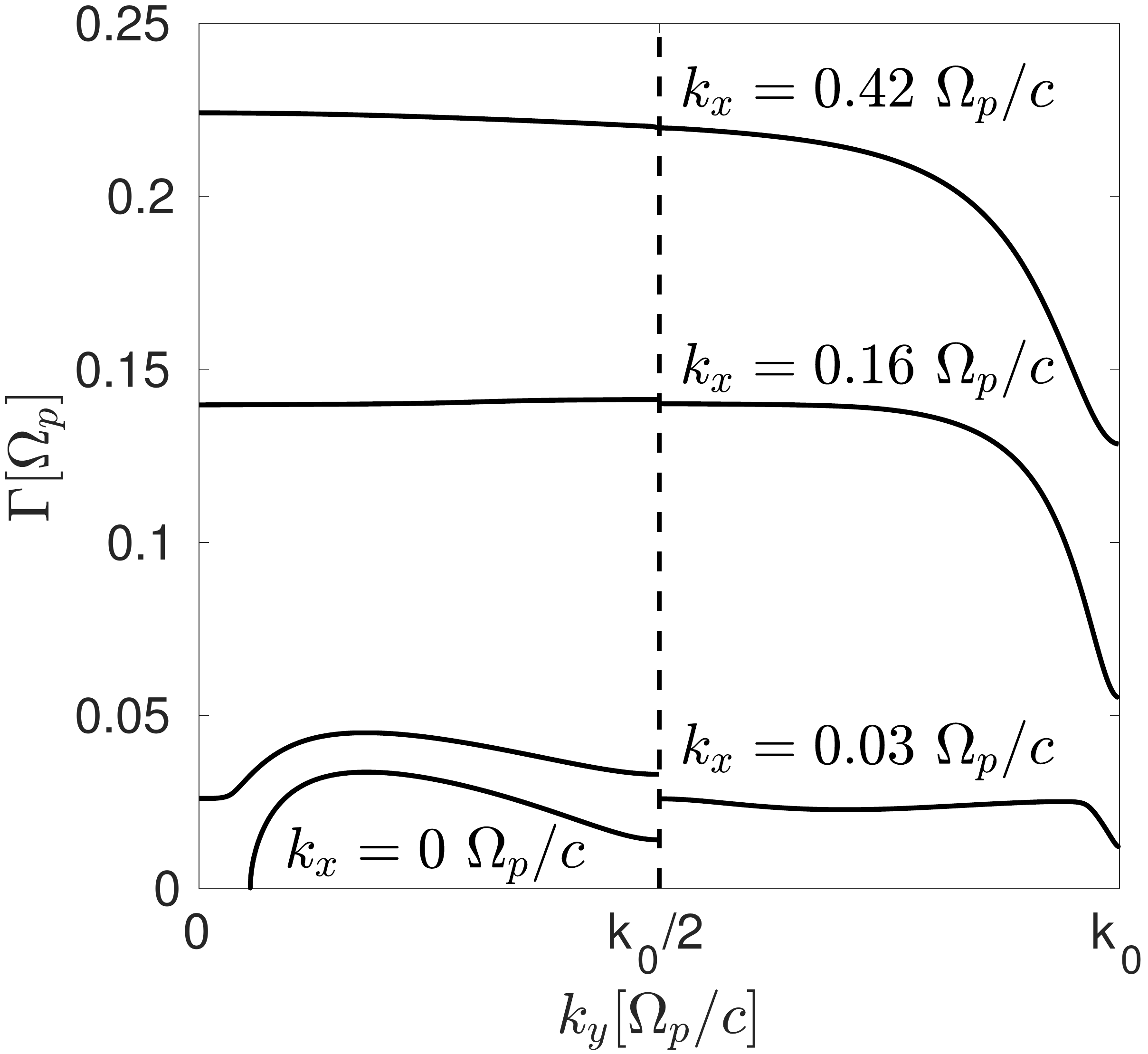}
	\caption{Instability growth rate ($\Gamma$) as a function of the characteristic Floquet exponent ($k_y$) for different values of $k_x$. 
		The stationary symmetric system is characterized by $T_0 = 1$, $\gamma_0 = 10$ and $\xi = 5$, yielding $\lambda_0=1.28\,c/\Omega_p$.}
	\label{fgr:disp_non_lin}
\end{figure}

We now perform a full 2D stability analysis of the equilibrium filamentary state by considering nonzero values of the longitudinal wavenumber $k_x$.
Since the two-beam system remains symmetric, the $\omega$ sampling space is still restricted to the imaginary axis.  In Fig.~\ref{fgr:disp_non_lin_tot},
we plot the $k_y$-maximized growth rate as a function of $k_x$ for various degrees of nonlinearity. As previously discussed, the dominant mode at
$\xi = 0.4$ is the purely tranverse ($k_x=0$) FMI. As we increase $\xi$, the growth rates globally decrease, yet the FMI modes near $k_x=0$ are
the most severely weakened. At $\xi = 2.5$, the growth rate curve exhibits a plateau at $\Gamma \simeq 0.22\,\Omega_p$ in the interval
$0 \le k_x \le 0.6\,\Omega_p/c$. On further raising $\xi$, the low-$k_x$ FMI modes are increasingly stabilized, while modes around
$k_x \gtrsim 0.4\,\Omega_p/c$ are only weakly diminished. At $\xi = 5$, this trend leads to the growth rate curve peaking at $k_{x,\rm max} \simeq 0.42\,\Omega_p$.

Figure~\ref{fgr:disp_non_lin} shows the $k_y$-dependence of $\Gamma$ for $\xi=5$ and different values of $k_x$. The $k_x=0$ curve is identical to that plotted in Fig.~\ref{fgr:disp_rel_sym}, as discussed above.
At $k_x=0.03\,\Omega_p/c$, the entire range $0 \le k_y \le k_0$ is destabilized, with finite and increased growth rates everywhere
(even at $k_y =0$ and $k_0$). A gap is still present at $k=k_0/2$ and the dominant mode remains located at a relatively low $k_y\simeq 0.8\,\Omega_p/c$. Further increasing $k_x$ strengthens all modes, but especially those with $k_y> k_0/2$, so that the gap at $k_y=k_0/2$ is progressively bridged. The $\Gamma$ curves then tend to form a quasi-plateau in the $0 \le k_y < k_0/2$ range, and drop to a finite value for $k_y\,\to\,k_0$.

\begin{figure}[t!]
  \centering
  \includegraphics[width=0.9\hsize]{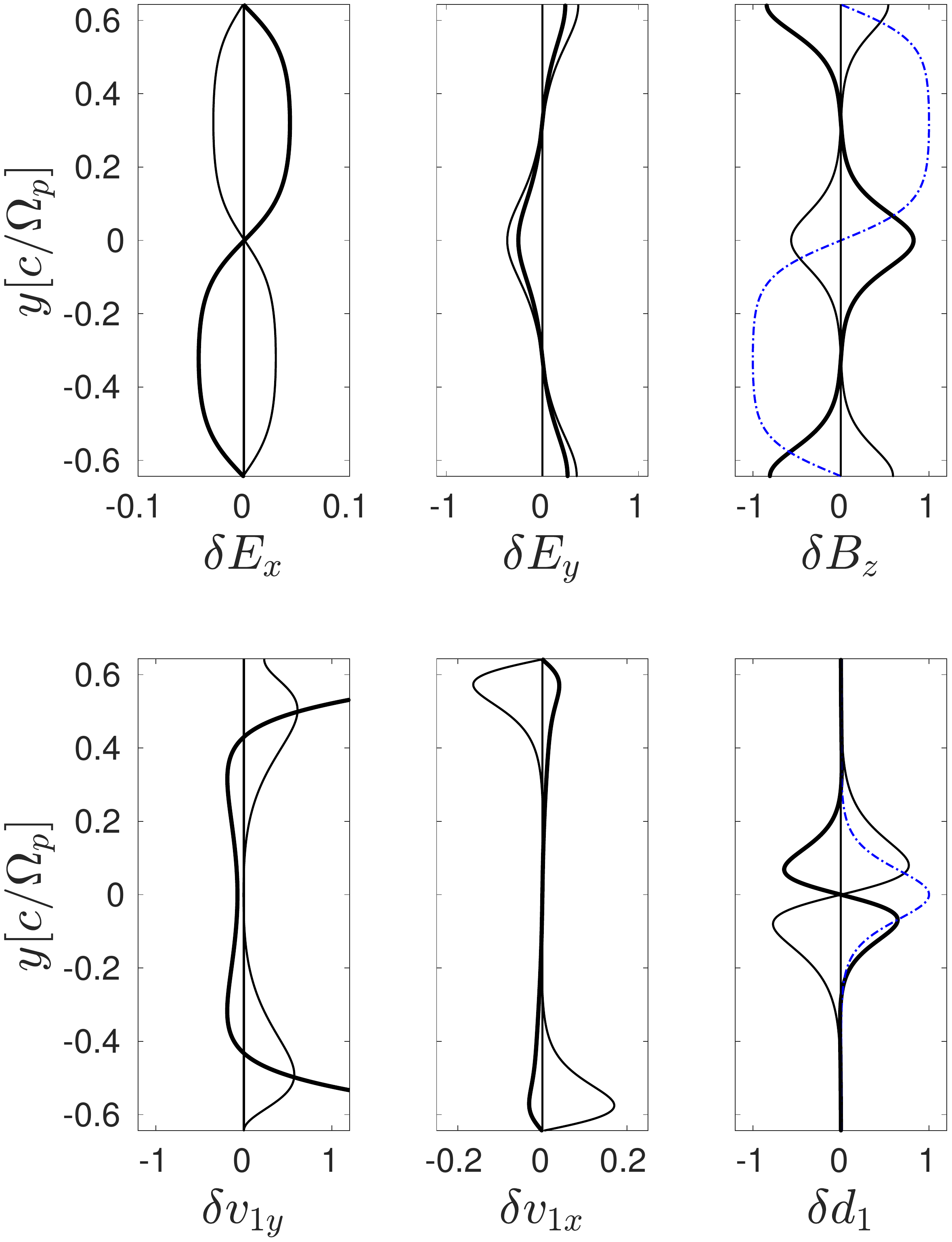}
  \caption{Transverse spatial structure of the eigenmode characterized by $\Gamma=0.22\,\Omega_p$, $k_y = 0$ and $k_x = 0.42\,\Omega_p/c$
  (same parameters as in Fig.~\ref{fgr:disp_non_lin_tot}). This eigenmode is the dominant one in Figs.~\ref{fgr:disp_non_lin_tot} and \ref{fgr:disp_non_lin}.
  The thick (resp. thin) solid lines plot the real (resp. imaginary) part of the eigenfunctions.  The perturbed quantities are, from left to right and top to bottom, $\delta E_x$, $\delta E_y$, $\delta B_z$, normalized to $\mathrm{max}_y \vert \delta B_z \vert$
  and $\delta v_{1x}$, $\delta v_{1y}$ and $ \delta d_1$, normalized to $\mathrm{max}_y \vert \delta d_1 \vert$. The thin dot-dashed lines
  plot the unperturbed quantities (normalized to their peak values).}
  \label{fgr:prof_TOP}
\end{figure}

\begin{figure}[t!]
  \centering
  \includegraphics[width=0.9\hsize]{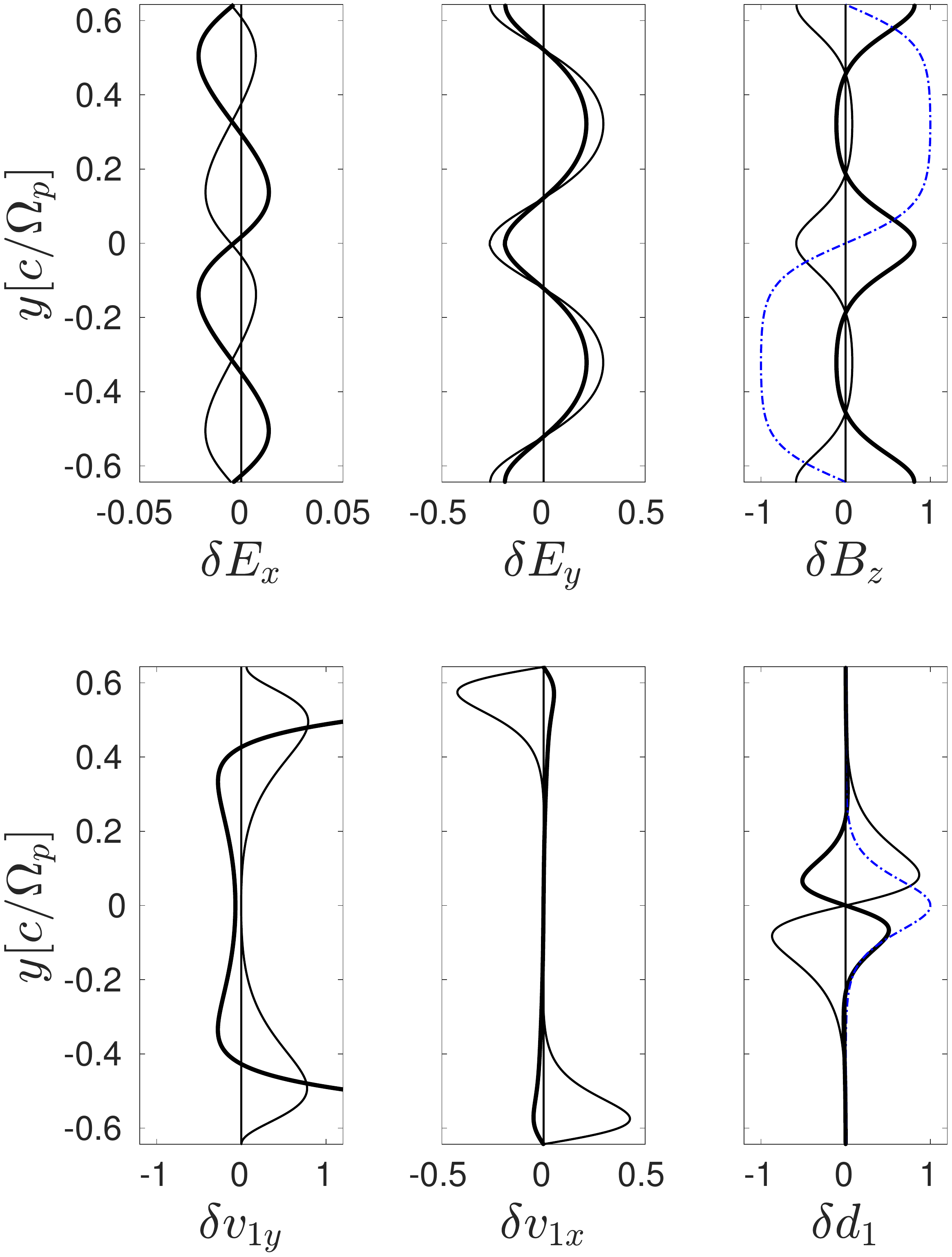}
  \caption{Transverse spatial structure of the eigenmode characterized by $\Gamma=0.13\,\Omega_p$, $k_y = k_0$ and $k_x = 0.42\,\Omega_p/c$
  (same parameters as in Fig.~\ref{fgr:disp_non_lin_tot}). The thick (resp. thin) solid lines plot the real (resp. imaginary) part of the eigenfunctions.
  The perturbed quantities are, from left to right and top to bottom, $\delta E_x$, $\delta E_y$, $\delta B_z$, normalized to $\mathrm{max}_y \vert \delta B_z \vert$
  and $\delta v_{1x}$, $\delta v_{1y}$ and $ \delta d_1$, normalized to $\mathrm{max}_y \vert \delta d_1 \vert$. The thin dot-dashed lines
  plot the unperturbed quantities (normalized to their peak values).}
  \label{fgr:prof_BOT}
\end{figure}

For $k_x=0.42\,\Omega_p/c$, the growth rates are bounded between $\Gamma = 0.22\,\Omega_p$ ($k_y \to 0$) and $\Gamma = 0.13\,\Omega_p$ ($k_y \to k_0$), while the discontinuity at $k_y=k_0/2$ is no longer visible. The eigenmodes associated with $k_y=0$ and $k_y=k_0$ have a minimal periodicity that is equal to,
or is a divisor of the fundamental period of the equilibrium system. Their transverse spatial structure is displayed in Figs.~\ref{fgr:prof_TOP} and 
\ref{fgr:prof_BOT}, respectively. We expect the intermediate eigenfunctions (with $0 < k_y < k_0$) to evolve smoothly between
these two limits. Each figure plots the $(\delta E_x, \delta E_y, \delta B_z)$ and $(\delta v_{1y}, \delta v_{1x}, d_1)$ components of the eigenmode,
respectively normalized to $\mathrm{max}_y \vert B_z \vert$ and $\mathrm{max}_y \vert d_1 \vert$. The thick and thin solid lines represent, respectively,
the real and imaginary parts of each quantity. The perturbed magnetic (resp. density) profile of both eigenmodes is even (resp. odd) with respect to the
filament center, which is indicative of DKI \cite{Zenitani_2007}. The two modes mainly differ in
the oscillation phase between opposite-current filaments. For the dominant mode, the magnetic perturbations at the center of two neighboring current
filaments (\emph{e.g.}, at $y=0$ and $y=0.64\,c/\Omega_p$) are of opposite sign, so that all filaments oscillate in phase. For the sub-dominant mode, 
adjacent filaments of opposite current see magnetic perturbations of the same sign, and hence they oscillate out of phase. As a consequence, the electromagnetic
profiles of the two modes have different periodicities: the dominant mode has the same minimal periodicity ($\lambda_0$) as the equilibrium system,
whereas the sub-dominant mode has a minimal periodicity of $\lambda_0/2$.

\begin{figure*}[t!]
  \begin{tabular}{c}
  \includegraphics[width=0.8\textwidth]{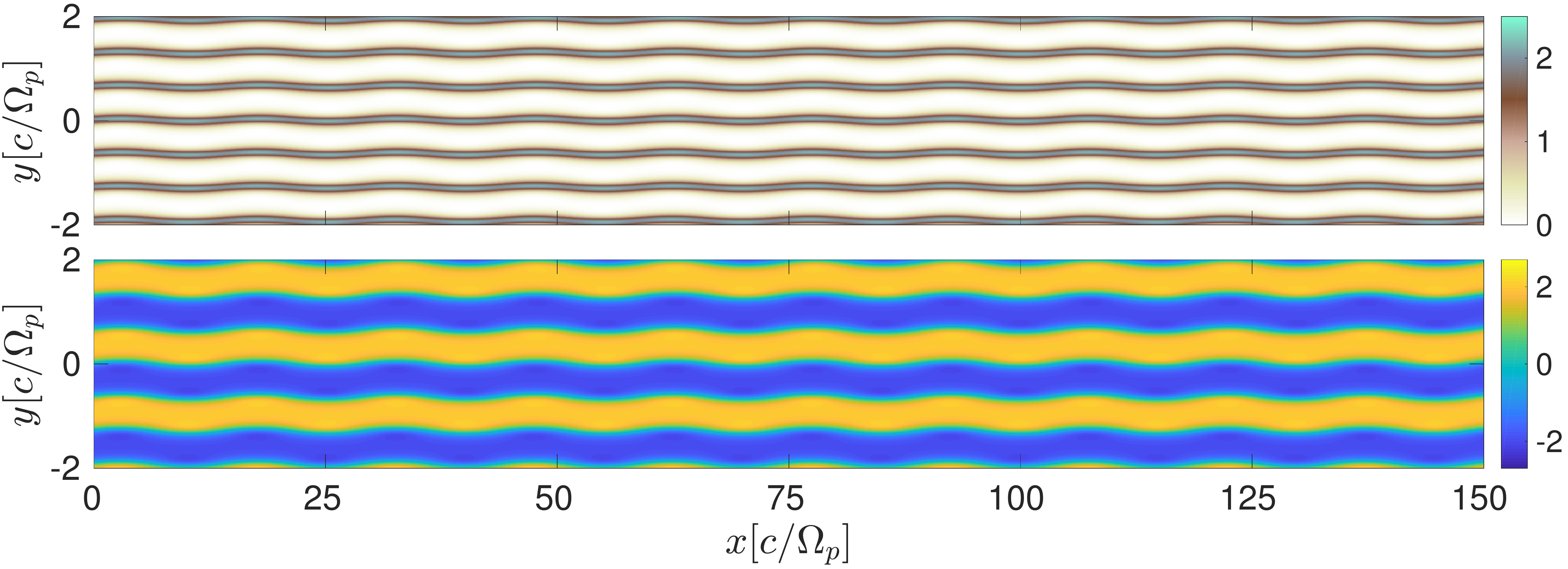}
  \end{tabular}
  \caption{2D spatial structure of the eigenmode characterized by $\Gamma=0.22\,\Omega_p$, $k_y = 0$ and $k_x = 0.42\,\Omega_p/c$. 
  The unperturbed symmetric system is defined by $T_0 = 1$, $\gamma_0 = 10$ and $\xi = 5$.
  Top panel: total particle density ($\sum_\alpha d_\alpha$). Bottom panel: transverse magnetic field ($B_z$).} 
  \label{fgr:floquet_top}
\end{figure*}

\begin{figure*}[t!]
  \begin{tabular}{c}
  \includegraphics[width=0.8\textwidth]{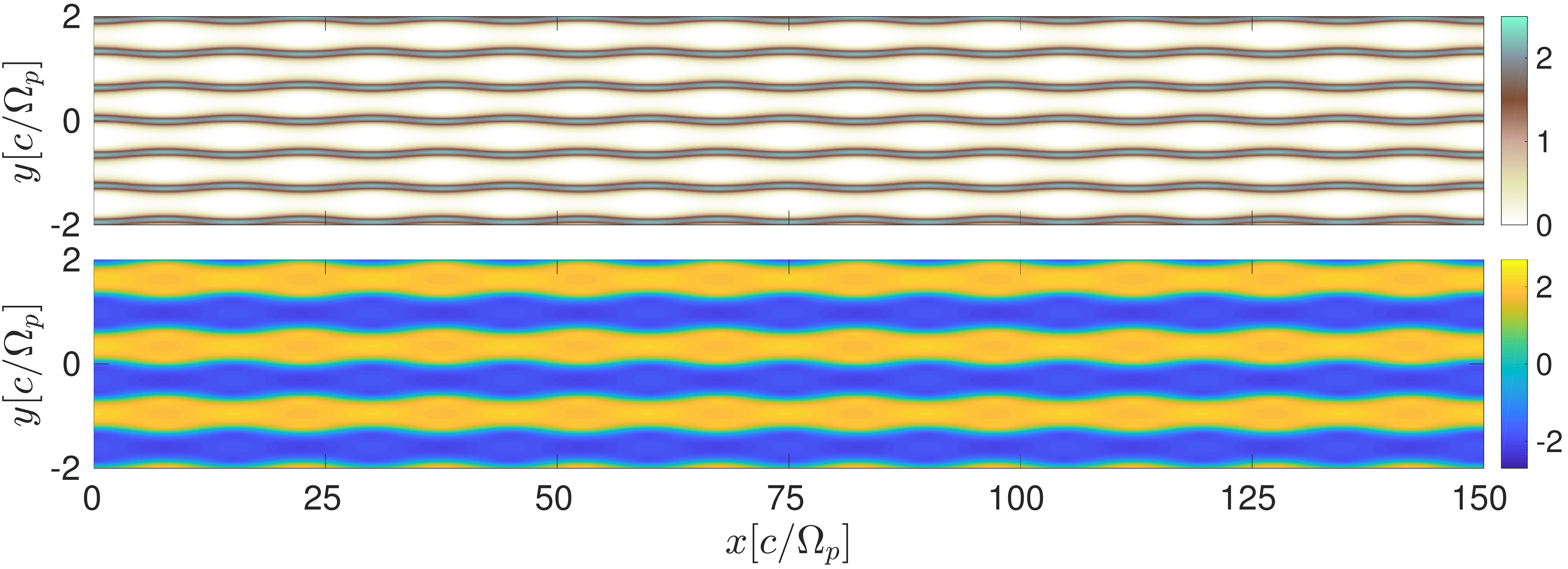}
  \end{tabular}
  \caption{2D spatial structure of the eigenmode characterized by $\Gamma=0.13\,\Omega_p$, $k_y = k_0$ and $k_x = 0.42\,\Omega_p/c$
  (same parameters as in Fig.~\ref{fgr:floquet_top}).
  Top panel: total particle density ($\sum_\alpha d_\alpha$). Bottom panel: transverse magnetic field ($B_z$).}
  \label{fgr:floquet_bot}
\end{figure*}

For completeness, we construct in Figs.~\ref{fgr:floquet_top} and \ref{fgr:floquet_bot} synthetic 2D images of the two eigenmodes by adding, for both
the total apparent density ($\sum_\alpha d_\alpha$) and the magnetic field ($B_z$), the zeroth and first order terms. The first order term is multiplied by a factor
large enough to clearly illustrate the instability pattern. As expected, the dominant mode gives rise to in-phase oscillations of all filaments, and 
therefore to similar kinked deformations of $\sum_\alpha d_\alpha$ and $B_z$ (Fig.~\ref{fgr:floquet_top}). For the sub-dominant mode (Fig.~\ref{fgr:floquet_bot}),
the out-of-phase oscillations of adjacent current filaments translate into sausage-type magnetic fluctuations.


\subsection{\label{sec:level2_strong_2}Transition from FMI to DKI}

We now derive an analytic formula for the transition between dominant filament-merging and drift-kink instabilities in a symmetric relativistic
filamentary system. To this goal, we seek an approximate expression for the relativistic DKI assuming negligible coupling effects between
neighboring filaments. Our calculation, detailed in Appendix~\ref{app:DKI}, draws upon the approaches used in
Refs.~\cite{Pritchett_1996, Daughton_1999, Zenitani_2007} in the non- or weakly relativistic regimes. Here we only summarize its
main steps and results.

\begin{figure}[t!]
  \centering
  \includegraphics[width=0.8\hsize]{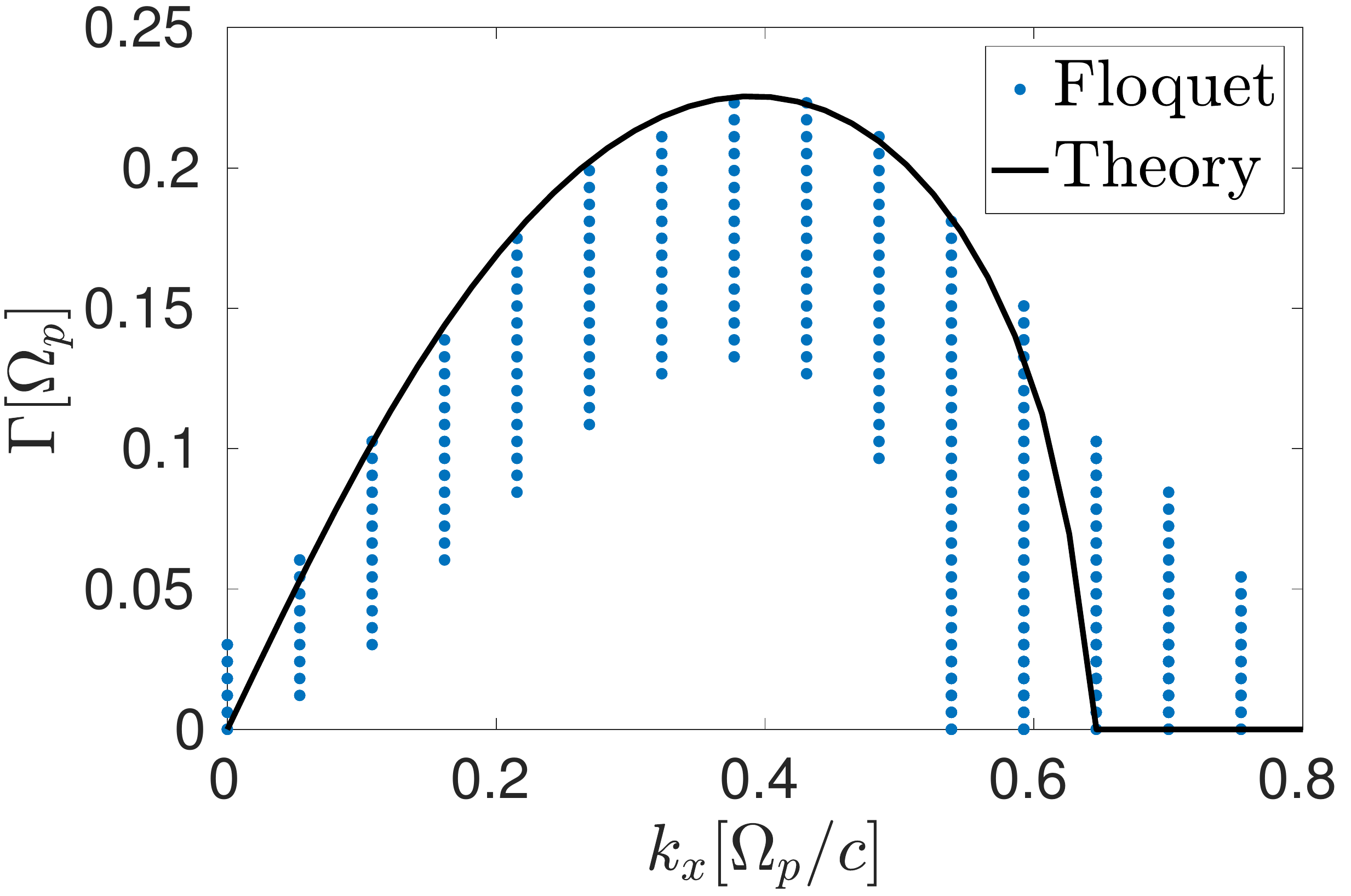}
  \caption{Growth rate of the relativistic drift kink instability as a function of $k_x$ for a two-beam filamentary system with
   $T_0 = 1$, $\gamma_0 = 10$, $a_0 = 1$ ($\xi = 5$) and $\Gamma_{\rm ad} = 4/3$. The results of the Floquet analysis are shown as
   blue points, while the solid black curve plots the analytic formula \eqref{eq:gamma_K_th} for an isolated current filament.}
  \label{fgr:kink_floquet}
\end{figure}

Let us first consider a highly relativistic current filament, for which we assume
\begin{equation}
  \bm{\beta} \cdot \delta \bm{\beta} = 0 \,,
\end{equation}
in agreement with the spatial profiles of Fig.~\ref{fgr:prof_TOP}. Furthermore, we assume that the pressure balance condition is satisfied,
\begin{equation}
  \frac{\mathbf{B} \cdot \delta{\mathbf{B}}}{4 \pi} + \delta{p}_+ + \delta{p}_- = 0 \,,
\end{equation}
where $(+,-)$ refers to the charge of the two species making up the current filament. This condition is compatible with the profiles obtained from the
Floquet theory. By substituting the above relations into the linearized two-fluid-Maxwell system, one can obtain the following expression for the DKI
growth rate ($\Gamma_\mathrm{DKI}$) in a relativistic Harris-type current sheet:
\begin{equation}\label{eq:gamma_K_th}
  h_0 \frac{\Gamma^2_\mathrm{DKI}}{\Omega_p^2} \simeq  \sqrt{1+4 h_0 \frac{k_x^2 c^2}{\Omega_p^2}} - 1 - h_0 \frac{k_x^2c^2}{\Omega_p^2} \,,
\end{equation}
where $h_0 = 1 + \frac{\Gamma_{\rm ad}}{\Gamma_{\rm ad} - 1} T_0$. This formula predicts the maximum growth rate for the DKI
\begin{equation}\label{eq:max_gamma_K_th}
 \Gamma_\mathrm{DKI, max} \simeq \frac{\Omega_p}{2 \sqrt{h_0}} \,,
\end{equation}
reached at $k_x = \sqrt{3/4h_0}\,\Omega_p/c$.

In Fig.~\ref{fgr:kink_floquet}, we compare Eq.~\eqref{eq:gamma_K_th} with the DKI growth rates computed numerically using the Floquet method
for a strongly nonlinear periodic system defined by $T_0 = 1$, $\gamma_0 = 10$, $a_0 = 0.5$ ($\xi=5$) and $\Gamma_\mathrm{ad} = 4/3$. 
For a given value of $k_x$, each blue point represents a solution to the Floquet problem. We observe that the analytic formula closely
reproduces the upper envelope of the numerical growth rates, thus proving that inter-filament effects are indeed negligible in the
(symmetric) strong-pinching regime.

In order to determine which instability (FMI or DKI) dominates a given symmetric system, one must compare the maximum growth rate of the DKI
[Eq.~\eqref{eq:max_gamma_K_th}] with that of the FMI. An accurate formula for the latter should, in principle, account for the inhomogeneity of the
unperturbed system, which is usually done via a truncated Floquet expansion \cite{Mishra_2014}. Here we opt for a much simpler approach, based
on the observation that the FMI behaves similarly to the CFI in the very weakly nonlinear (or quasi homogeneous) limit (see Fig.~\ref{fgr:disp_rel_sym}).
In Appendix~\ref{app:CFI}, the following expression for the ultra-relativistic limit of the maximum growth rate of the CFI is obtained
\begin{equation}\label{eq:max_gamma_W_th}
  \Gamma_\mathrm{CFI,max} \simeq \frac{2 \Omega_p}{\sqrt{h_0}} \exp (-\xi/2) \,.
\end{equation}
Since the nonlinearity of the filaments tends to quell the FMI, this expression represents an upper bound value of the FMI growth rate.

According to Eqs.~\eqref{eq:max_gamma_K_th} and \eqref{eq:max_gamma_W_th}, the DKI is predicted to prevail over the FMI when the
nonlinearity parameter exceeds the approximate threshold value 
\begin{equation}
  \xi_\mathrm{th} \simeq 4\,\log 2 \simeq 2.7 \,.
\end{equation}
This analytic criterion for the FMI-DKI transition is in close agreement with our numerical results, which show a transition at $\xi \simeq 2.5$.

\subsection{\label{sec:level2_strong_3}2D PIC simulation}

We now present the results of a 2D3V \textsc{calder} simulation using the parameters considered in Sec.~\ref{sec:level2_strong_1} ($T_0 = 1$, $\gamma_0 = 10$, $a_0 = 0.5$).
The domain size is $25000\,\Delta x \times 1000\,\Delta y$ with a discretization $\Delta x = \Delta y =\lambda_0/200 = 0.0064 \,c/\Omega_p$. Each pinched
filament extends over about 15 cells. These parameters allow us to resolve the $(k_x,k_y$) Fourier space in the range $[0.04, 976] \times [0.976,976]\,(\Omega_p/c)^2$.
Each cell initially contains 10 macro-particles per species. The Maxwell equations are solved using the Cole-Karkkainen scheme \cite{Cole_1997a, Cole_2002, Karkkainen_2006}, which enables us to use a large time step,
$\Delta t = 0.99\, \Delta x/c$. As usual, we apply periodic boundary  conditions to both fields and particles. 

\begin{figure}[t!]
  \centering
  \includegraphics[width=0.8\hsize]{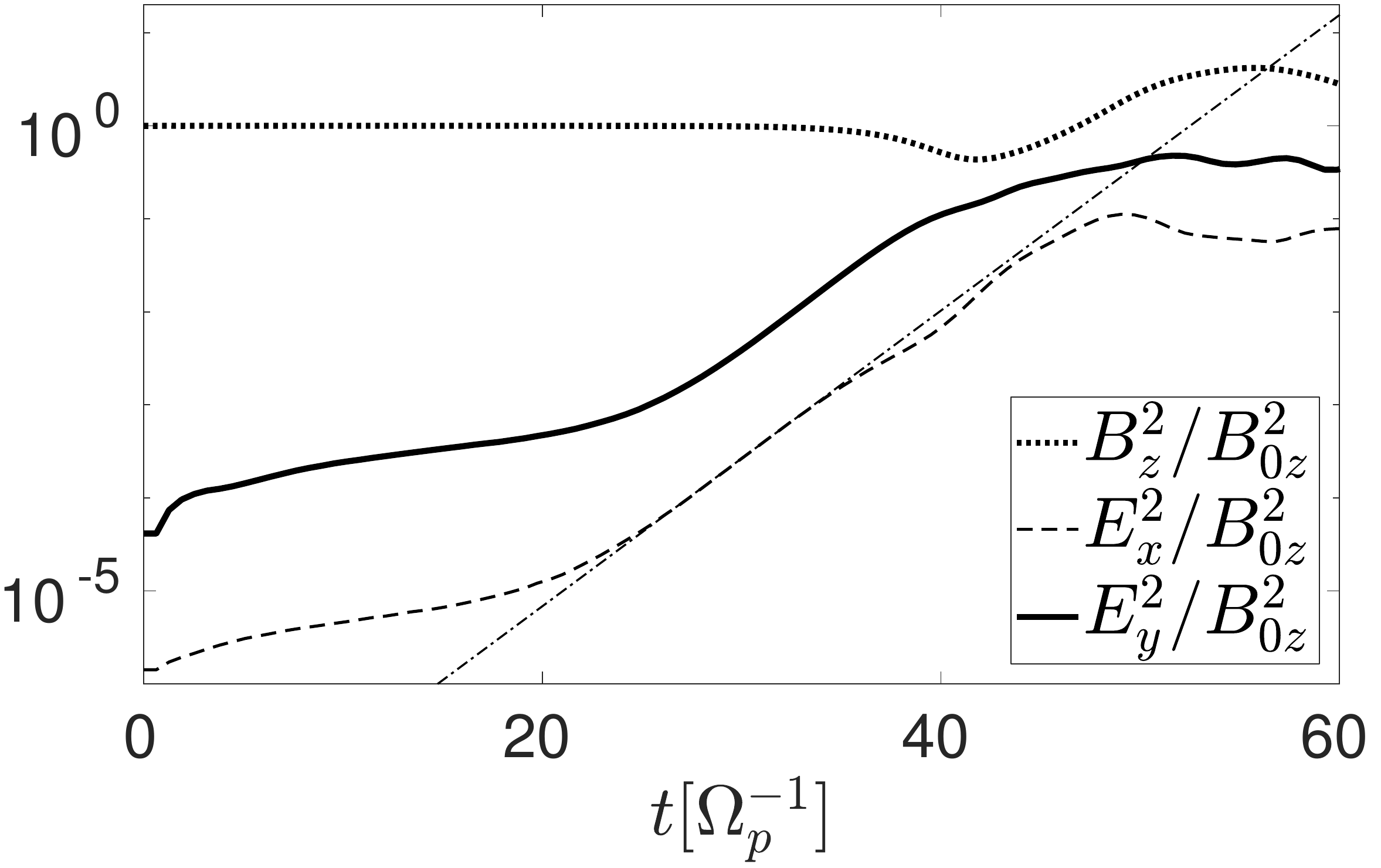}
  \caption{2D PIC simulation with the parameters $T_0=1$, $\gamma_0=10$ and $a_0=0.5$: time evolution of the electromagnetic energies (spatially integrated
  and normalized to the initial magnetic energy). The $B_z$ energy is plotted as a dotted line, the $E_x$ energy as a dashed line and the $E_y$ energy as a
  thick solid line. The growth rate, $\Gamma_\mathrm{PIC} \simeq 0.18\,\Omega_p$ (thin dashed-dotted line), is measured from the $E_y$ curve over the time
  interval $25 \le t \le 35\,\Omega_p^{-1}$.}
  \label{fgr:Ey_strong}
\end{figure}

\begin{figure*}[t]
  \begin{tabular}{c}
  \includegraphics[width=0.8\hsize]{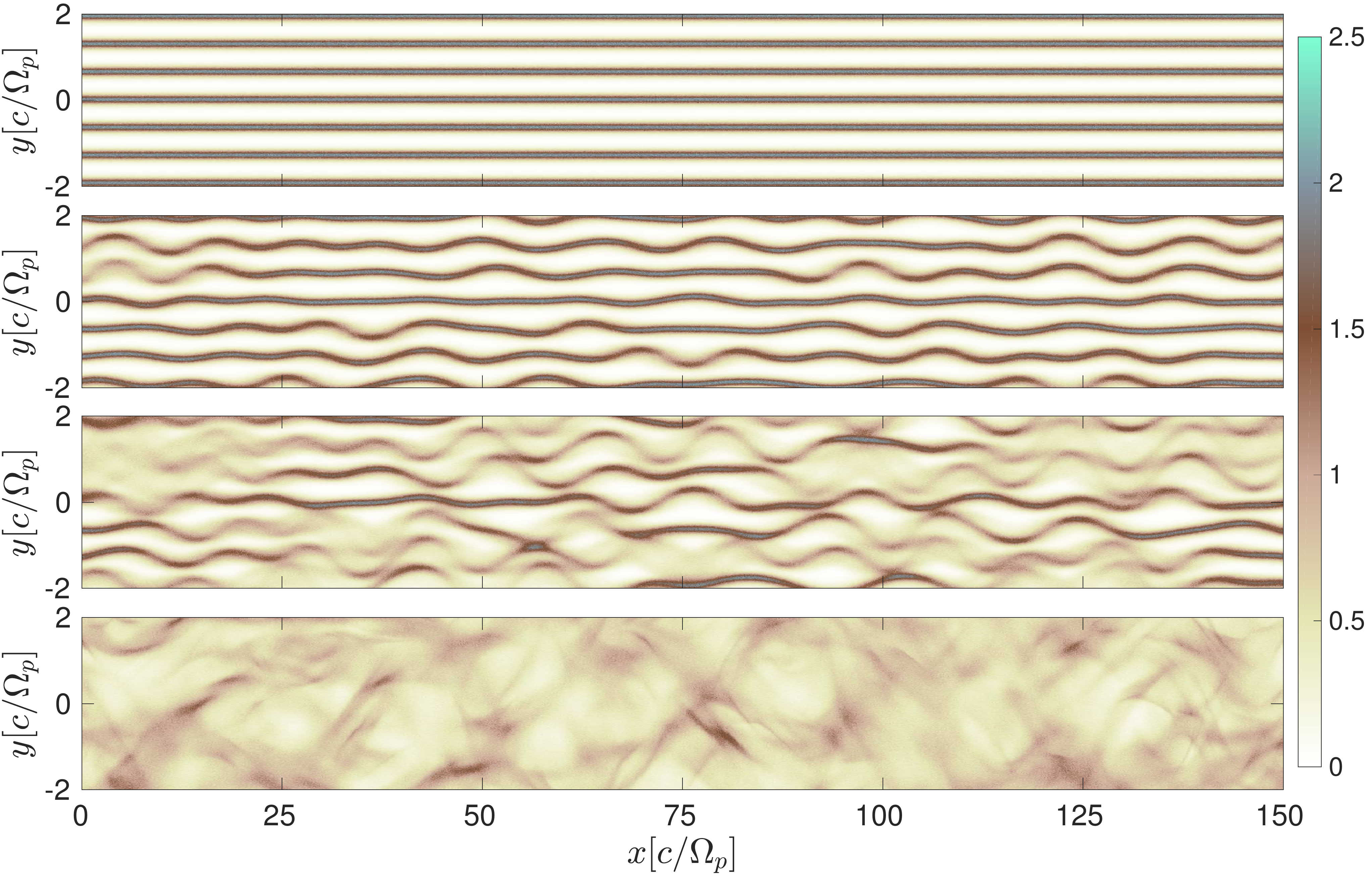}
  \end{tabular}
  \caption{Simulated profiles of the total particle density (normalized to the initial peak density of the positrons) for the parameters of
  Fig.~\ref{fgr:Ey_strong} at different times. From top to bottom: $t = 7,\ 34,\ 39,\ 50\ \Omega_p^{-1}$.}
  \label{fgr:D1_strong}
\end{figure*}

\begin{figure*}
  \begin{tabular}{c}
  \includegraphics[width=0.8\hsize]{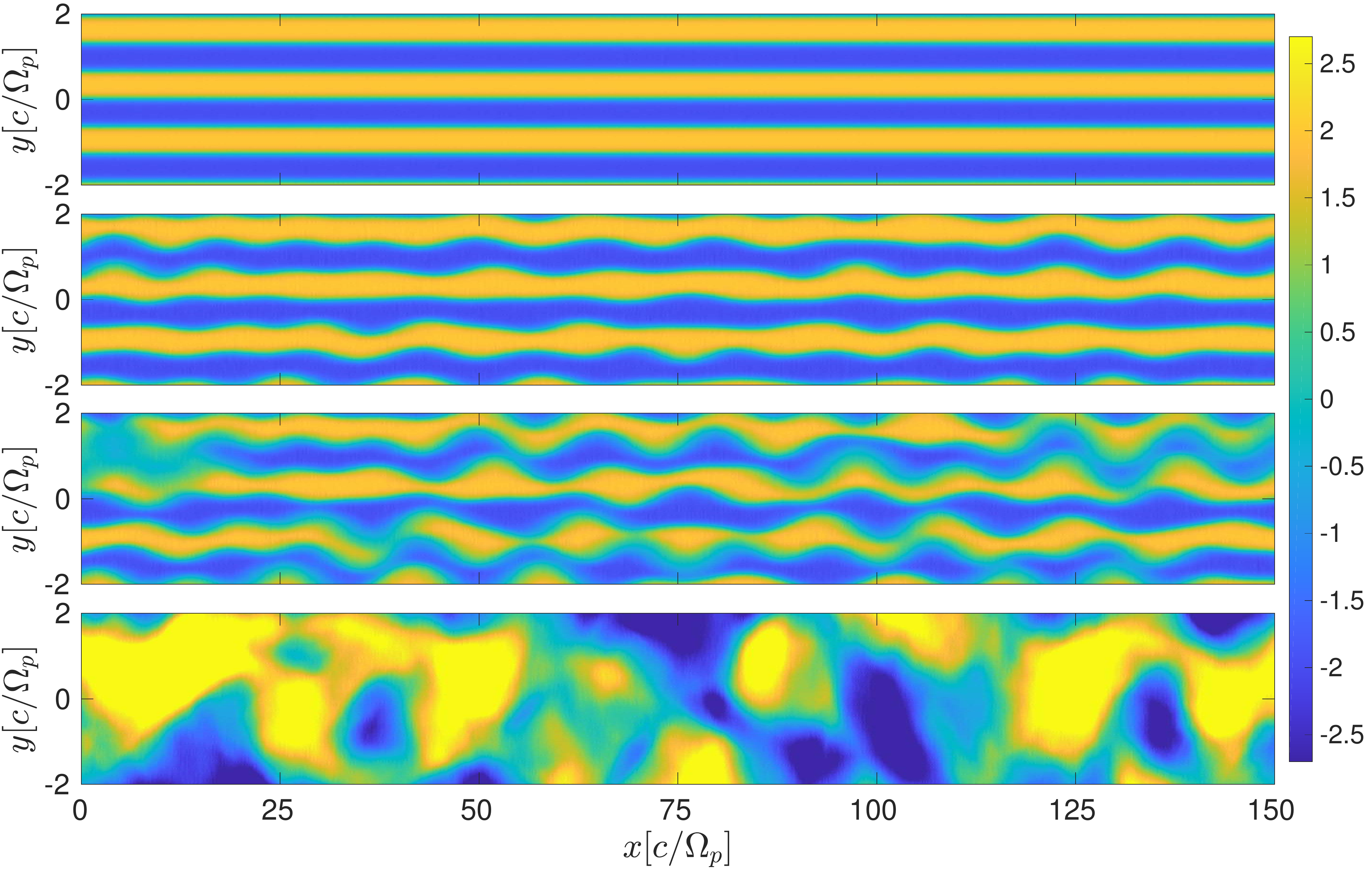}
  \end{tabular}
  \caption{Simulated magnetic-field ($B_z$, normalized to $m\Omega_p/e$) profile for the parameters of Fig.~\ref{fgr:Ey_strong} at different times.
  From top to bottom: $t = 7,\ 34,\ 39,\ 50\ \Omega_p^{-1}$.}
  \label{fgr:Bz_strong}
\end{figure*}

Figure~\ref{fgr:Ey_strong} plots the time history of the electromagnetic energies, while Figs.~\ref{fgr:D1_strong} and \ref{fgr:Bz_strong} display, respectively,
the maps of the magnetic field ($B_z$) and of the cold-beam-positron density ($d_1$) at times $t = 7,\ 34,\ 39$ and $50\,\Omega_p^{-1}$.
According to Fig.~\ref{fgr:Ey_strong}, the linear phase of the primary instability lasts until $t \simeq 40\,\Omega_p^{-1}$. During this period,
the current filaments develop kink oscillations (Fig.~\ref{fgr:D1_strong}). However, due to the broad unstable spectrum revealed in Fig.~\ref{fgr:disp_non_lin_tot},
they oscillate with a range of wavelengths and phases. While coherent motion between adjacent filaments can be seen locally, no single-mode pattern
clearly emerges at the end of the linear phase. In like manner, the $B_z$ field distribution (Fig.~\ref{fgr:Bz_strong}) exhibits a mix of kink- and
sausage-type perturbations instead of the globally coherent pattern displayed in Fig.~\ref{fgr:floquet_top}. 

As for the 1D FMI (Fig.~\ref{fgr:fft}), the 2D instability is first evidenced by exponentially growing electric-field energies (Fig.~\ref{fgr:Ey_strong}).
An effective growth rate $\Gamma_\mathrm{PIC} \simeq 0.18\,\Omega_p$ is measured in the linear instability phase ($25 \le t \le 35\,\Omega_p^{-1}$),
comparable with the maximum theoretical value, $\Gamma_\mathrm{max} \simeq 0.22\,\Omega_p$. In agreement with the predicted structure of the
dominant mode (Fig.~\ref{fgr:prof_TOP}), the $E_y$ energy is about an order of magnitude larger than the $E_x$ energy in the exponential phase. This feature
starkly contrasts with the behavior of the 1D FMI treated in Sec.~\ref{sec:level2_weak_2} (in the symmetric two-beam regime), for which the $E_y$ energy hardly grows
(Fig.~\ref{fgr:fft}). Another major difference with the FMI is that the saturation of the DKI ($35 \lesssim t \lesssim 40\,\Omega_p^{-1}$)
goes along with a sudden drop in the total magnetic energy (to $\sim 43\%$ of its initial value). This magnetic-field dissipation is an expected effect of the
DKI, which causes efficient particle heating \cite{Zenitani_2007}.

Once the initially strongly pinched filaments have been significantly distorted and smoothed out (see Figs.~\ref{fgr:D1_strong} and \ref{fgr:Bz_strong}
at $t=39\,\Omega_p^{-1}$), a secondary instability of the CFI type is triggered by the residual momentum anisotropy of the plasma: this causes the magnetic energy
to rebound again at $t \gtrsim 40 \,\Omega_p^{-1}$, and rise until $t\simeq 50\,\Omega_p^{-1}$, at which time it seems to saturate. The bottom panel of
Fig.~\ref{fgr:D1_strong} ($t = 50 \,\Omega_p^{-1}$) indicates that the plasma homogenization is then almost complete. Figure~\ref{fgr:Bz_strong}, however,
shows that the transverse wavelength of the secondary CFI is typically the transverse domain size. A more precise description of the late-time nonlinear evolution
of the system (which exceeds the scope of the present study) would therefore necessitate a larger simulation box.


\section{\label{sec:level1_asym}2D instability of asymmetric filaments}

\begin{figure}[t!]
  \centering
  \includegraphics[width=0.8\hsize]{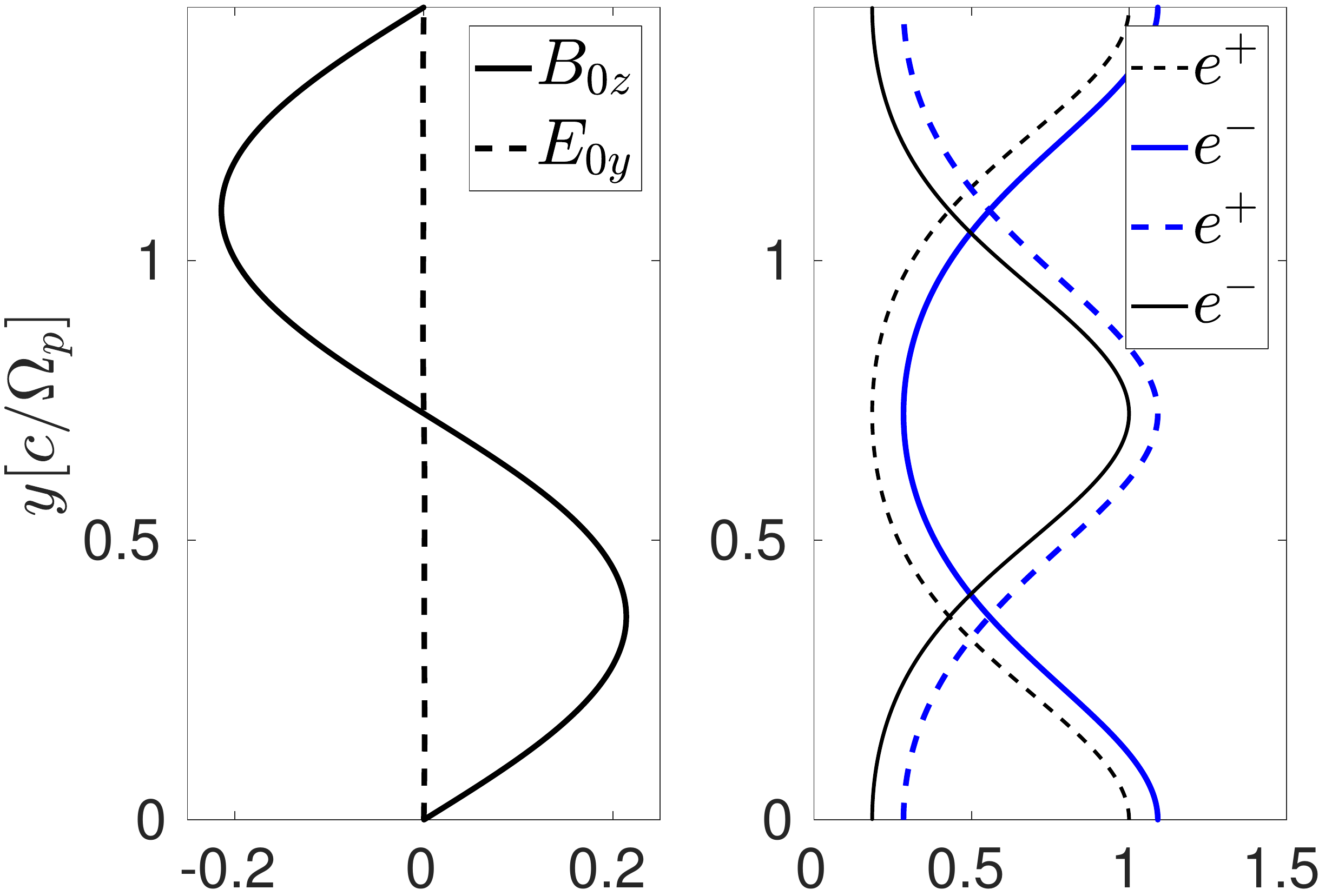}
  \caption{Stationary solution to the fluid-Maxwell equations in an $e^-e^+$ system composed of a hot beam ($T_{b0} = 1$, $\beta_{b0}= -0.9971$)
  and a background cold plasma ($T_{p0}=0.01$, $\beta_{p0} = 0.1672$, $N_p/N_b=10$). The vector potential maximum is $a_0=0.05$, resulting
  in a wavelength $\lambda_0=1.45\,c/\Omega_p$.
  Left panel: $B_{0z}(y)$ (solid line) and $E_{0y}(y)$ (dashed line), normalized to $\mathrm{max}_y B_{0z}$.
  Right panel: Density of the electrons (solid lines) and positrons (dashed lines) in the hot (blue) and cold (black) beams,
  normalized to the maximum cold-beam electron density.}
  \label{fgr:p_prof_asym}
\end{figure}

Finally, we address the interaction of a cold background plasma (indexed by `$p$') with a counterstreaming hot beam (indexed by `$b$'), both composed of
electrons and positrons. In asymmetric configurations, $\omega$ acquires a real part, which renders the Fourier space sampling complex. Moreover,
not all frames are well defined for searching a purely real longitudinal wave number $k_x$. Here, this frame is chosen to be the so-called Weibel frame
\cite{Pelletier_2018}, in which the electrostatic field of the stationary state vanishes. In the linear limit, this frame can be determined by setting
$\phi_0 = 0$ in Eq.~\eqref{eq:phi_stat}, which gives the following relation between the plasma species
\begin{equation} \label{eq:NL_Weibel_frame}
  \sum\limits_\alpha \frac{N_\alpha \beta_{\alpha 0} \gamma_{\alpha 0}^2}{T_\alpha}  = 0 \,.
\end{equation}
When $N_p/N_b \gg 1$, the Weibel frame tends to the rest frame of the cold background plasma \cite{Pelletier_2018}. In the following, we focus
on a particular configuration defined by $T_{p0}=0.01$, $T_{b0}=1$, $\beta_{b0} = -0.9971$ ($\gamma_{b0}= 13$), $\beta_{p0} = 0.1672$, $\Gamma_{\rm ad, p}=5/3$,
$\Gamma_{\rm ad, b} = 4/3$ and $N_p/N_b = 10$.  The vector potential maximum is set to $a_0 = 0.05$, yielding $\xi_p \simeq 0.85$ and $\xi_b \simeq 0.66$,
and hence moderately pinched current filaments, as confirmed by the equilibrium spatial profiles plotted in Fig.~\ref{fgr:p_prof_asym}. The lab-frame is chosen
such that  the plasma parameters fulfill Eq.~\eqref{eq:NL_Weibel_frame}, leading to $\vert E_{0y} \vert \ll \vert B_{0z} \vert$ (see Fig.~\ref{fgr:p_prof_asym}). 
The stationary system has a wavelength $\lambda_0 \simeq 1.45\,c/\Omega_p$.

\subsection{\label{sec:level2_asym_1}Floquet analysis}

\begin{figure}[t!]
  \centering
  \includegraphics[width=0.8\hsize]{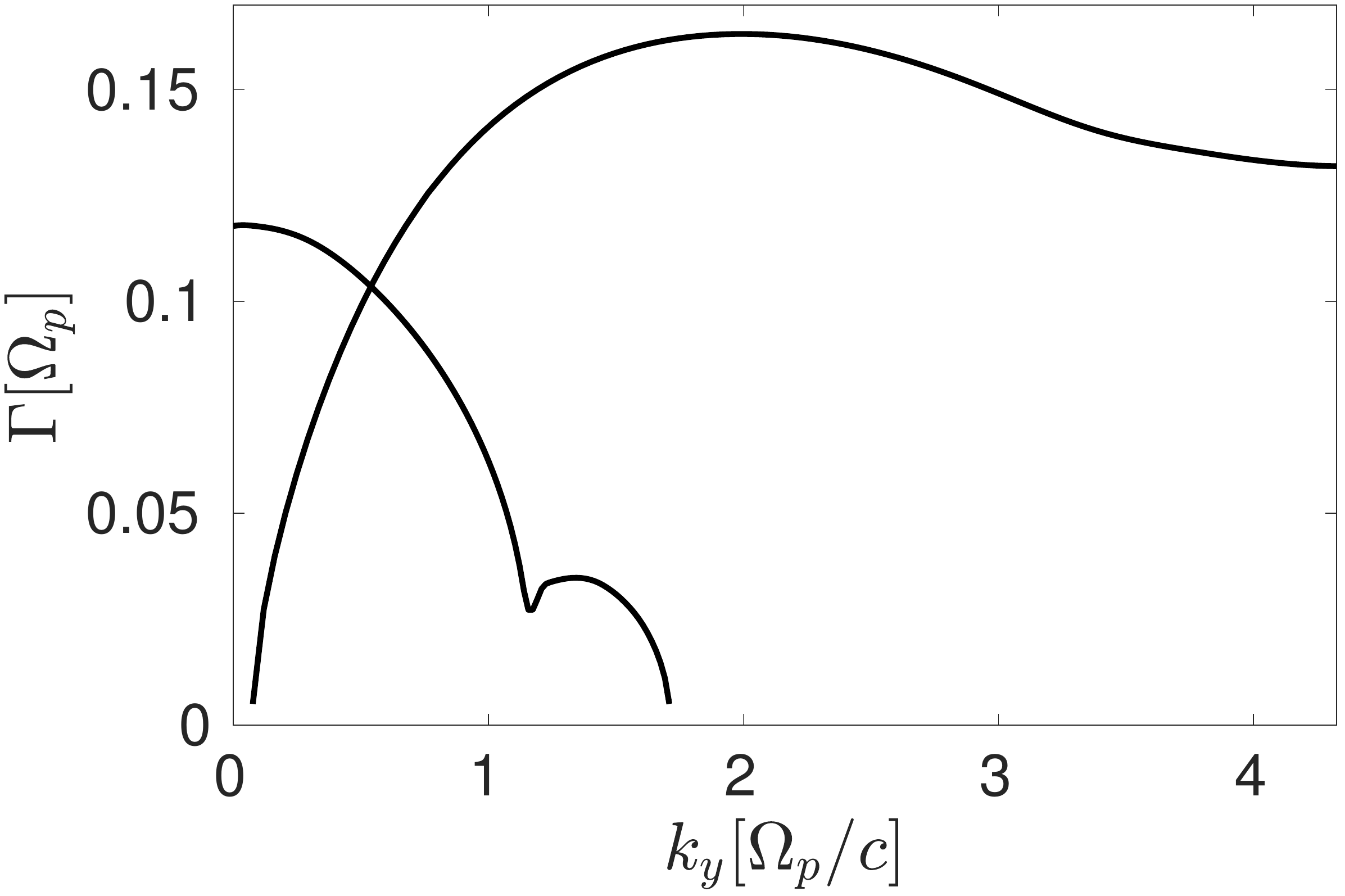}
  \caption{Instability growth rate ($\Gamma$) as a function of the Floquet exponent ($k_y$) at $k_x=0.95\,\Omega_p/c$ (corresponding to the fastest-growing
  mode). The unperturbed system is asymmetric with the parameters of Fig.~\ref{fgr:p_prof_asym}. The fundamental wavenumber is $k_0\,\equiv\,2\pi/\lambda_0\,=\,4.33 \,\Omega_p/c$.}
  \label{fgr:disp_asym_fl}
\end{figure}

The Floquet analysis is carried out by setting, for each real value of $k_x$, an imaginary value ($\Gamma$) for $\omega$, and varying its real value ($\omega_r$)
in order to find a vanishing value of $\Im k_y$. The roots are approached using a bisection method and computing the local derivative of $\Im k_y$.
Figure~\ref{fgr:disp_asym_fl} plots the $k_y$ dependence of $\Gamma$ for the longitudinal wavenumber ($k_x \simeq 0.95\,\Omega_p/c$) associated
with the fastest-growing mode. In contrast to the symmetric configuration, two unstable branches are found: the upper one extends over the full fundamental zone
($0\le k_y \le k_0 \simeq 4.3\,\Omega_p/c$), while the lower one is restricted to the range $0\le k_y \le 1.7\,\Omega_p/c$. The dominant mode pertains to
the upper branch, and is characterized by a Floquet exponent $k_y \simeq 2.0\,\Omega_p/c$, a growth rate $\Gamma \simeq 0.16\,\Omega_p$ and a real frequency
$\omega_r \simeq -0.75\,\Omega_p$ (Fig.~\ref{fgr:disp_asym_fl}). This mode differs significantly from the one governing the system in the homogeneous limit,
which is characterized by $k_x \simeq 1.3\,\Omega_p/c$, $k_y \simeq 3.2\,\Omega_p/c$, $\Gamma \simeq 0.25\,\Omega_p$ and $\omega_r \simeq -1.1 \,\Omega_p$.

\begin{figure*}[!]
  \begin{tabular}{c}
  \includegraphics[width=0.8\hsize]{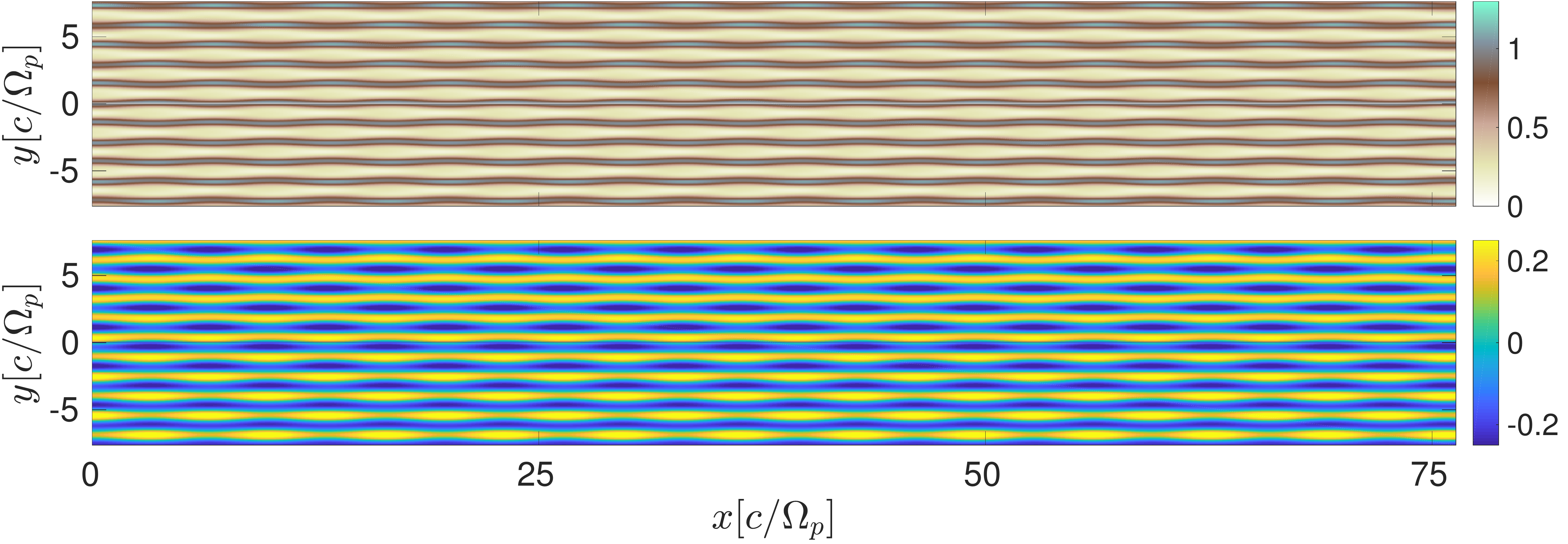}
  \end{tabular}
  \caption{2D spatial structure of the dominant eigenmodes ($k_y=\pm 2.0\,\Omega_p/c$) of the upper unstable branch shown in Fig.~\ref{fgr:disp_asym_fl} normalized to the same amplitude.
  Top panel:  apparent density of the cold-beam positrons ($d_1$). Bottom panel: transverse magnetic field ($B_z$). For each quantity, the zeroth
  and first order terms are added up.}
  \label{fgr:floquet_asym_dom}
\end{figure*}

Figure~\ref{fgr:floquet_asym_dom} displays the 2D structure of the dominant instability, as reflected by the apparent density of the cold-beam positrons ($d_1$)
and the transverse magnetic field ($B_z$). For both quantities, the zeroth and first order terms are added up. The first order term is taken to be the
sum of the eigenfunctions with $k_y = \pm 2.0\,\Omega_p/c$ and $k_x=0.95\,\Omega_p/c$ , which grow at the same rate and are initalized with the same amplitude.
The structure that we observe differs from that found in the symmetric case (see Figs.~\ref{fgr:prof_TOP} and \ref{fgr:prof_BOT}): the filaments are subject to a mix of kink and
bunching instabilities, with adjacent filaments of same species oscillating in opposite phase.

Figure~\ref{fgr:disp_asym_fl} shows that the modes of the upper unstable branch in the range $1 \lesssim k_y \lesssim k_0$ share similar
growth rates ($\Gamma \sim 0.13-0.16\,\Omega_p$). In comparison, the lower branch presents a more localized maximum at $k_y=0$. The spatial structure of this
sub-dominant mode corresponds to a bunching instability, as is the case for all lower-branch modes.


\subsection{\label{sec:level2_asym_2}2D PIC simulation}

\begin{figure*}[t!]
  \includegraphics[width=0.8\hsize]{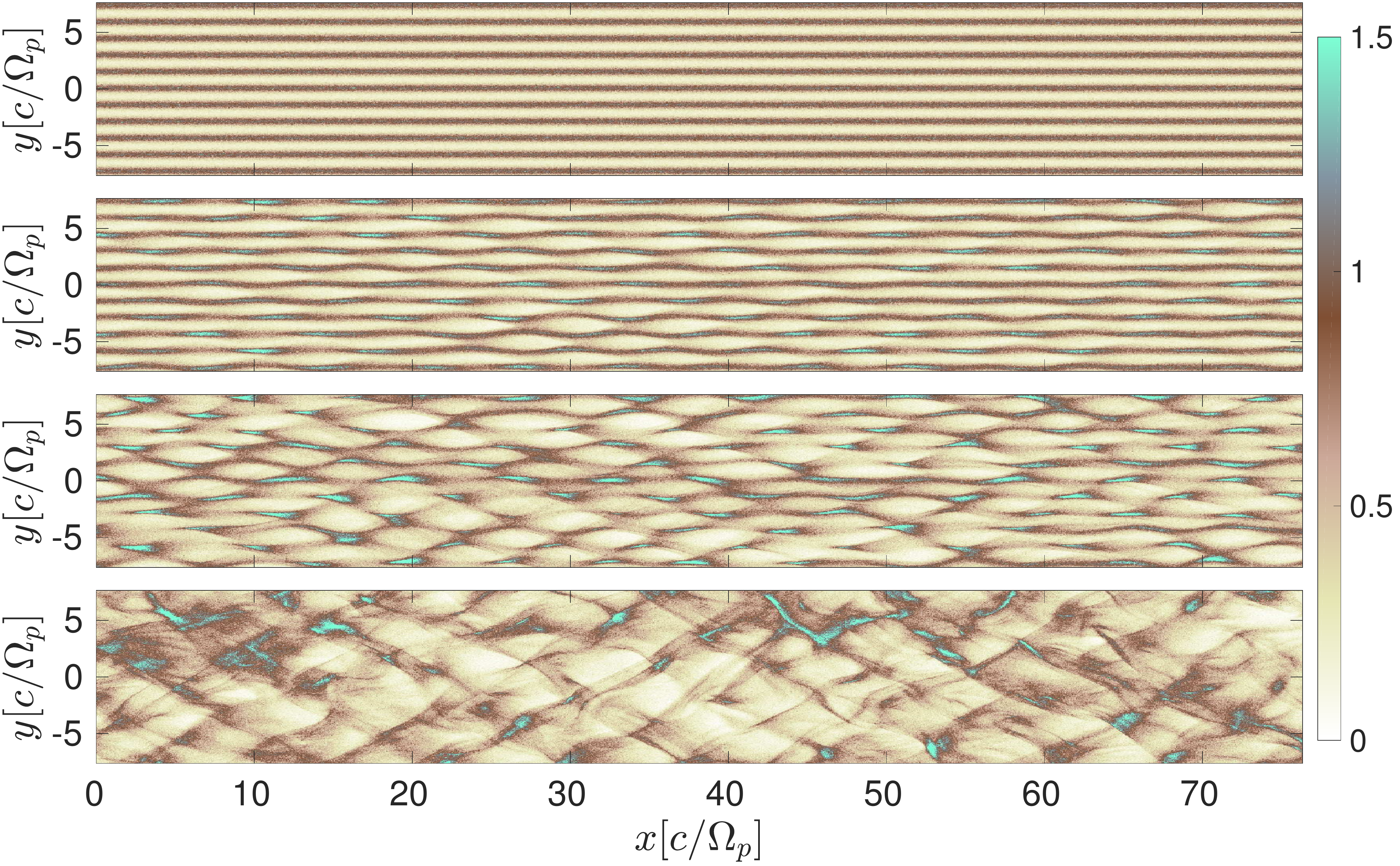}
  \caption{2D PIC simulation with the parameters of Fig.~\ref{fgr:p_prof_asym}: cold-beam positron density ($d_1$, normalized to its initial peak value) at
  different times. From top to bottom: $t = 3,\ 32,\ 41,\ 58\ \Omega_p^{-1}$.}
  \label{fgr:D1}
\end{figure*}

We have performed a 2D3V PIC simulation with the same initial parameters as the asymmetric Floquet problem. The numerical setup is that used in
Sec.~\ref{sec:level2_strong_3}, except that we now set $\Delta x = \Delta y = 0.0292\,c/\Omega_p$. The time history of the electromagnetic energies is plotted
in Fig.~\ref{fgr:D1}. From the evolution of the $E_x$ energy, the primary instability is estimated to grow at a rate $\Gamma_\mathrm{PIC} \simeq 0.14\,\Omega_p$,
close to the theoretical prediction ($\Gamma_\mathrm{max} \simeq 0.16\,\Omega_p$). During this early phase, the total $B_z$ energy remains essentially constant. 
The exponential rise in the $E_x$ energy comes to an end at $t \simeq 40\,\Omega_p^{-1}$, at which time a new, slower-growing instability kicks in,
giving rise to concomitant increases in the $E_y$ and $B_z$ energies.

\begin{figure}[t!]
  \includegraphics[width=\hsize]{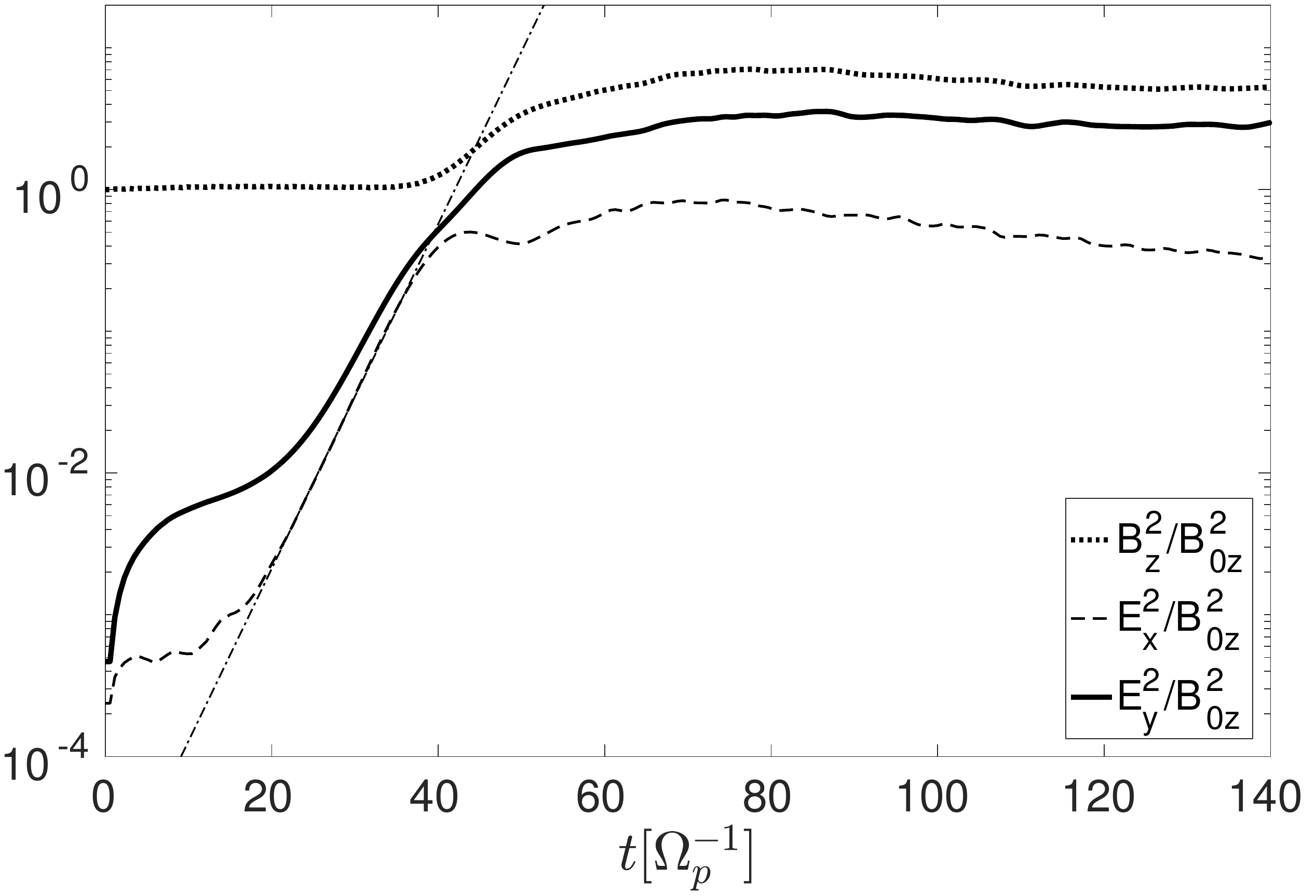}
  \caption{2D PIC simulation with the parameters of Fig.~\ref{fgr:p_prof_asym}: time evolution of the electromagnetic energies (spatially integrated
  and normalized to the initial magnetic energy). The $B_z$ energy is plotted as a dotted line, the $E_x$ energy as a dashed line and the $E_y$ energy as a
  thick solid line. The growth rate, $\Gamma_\mathrm{PIC} \simeq 0.14\,\Omega_p$ (thin dashed-dotted line), is measured from the $E_x$ curve over the time
  interval $20 \le t \le 35\,\Omega_p^{-1}$.}
  \label{fgr:disp_asym}
\end{figure}

\begin{figure*}[t!]
  \includegraphics[width=0.8\hsize]{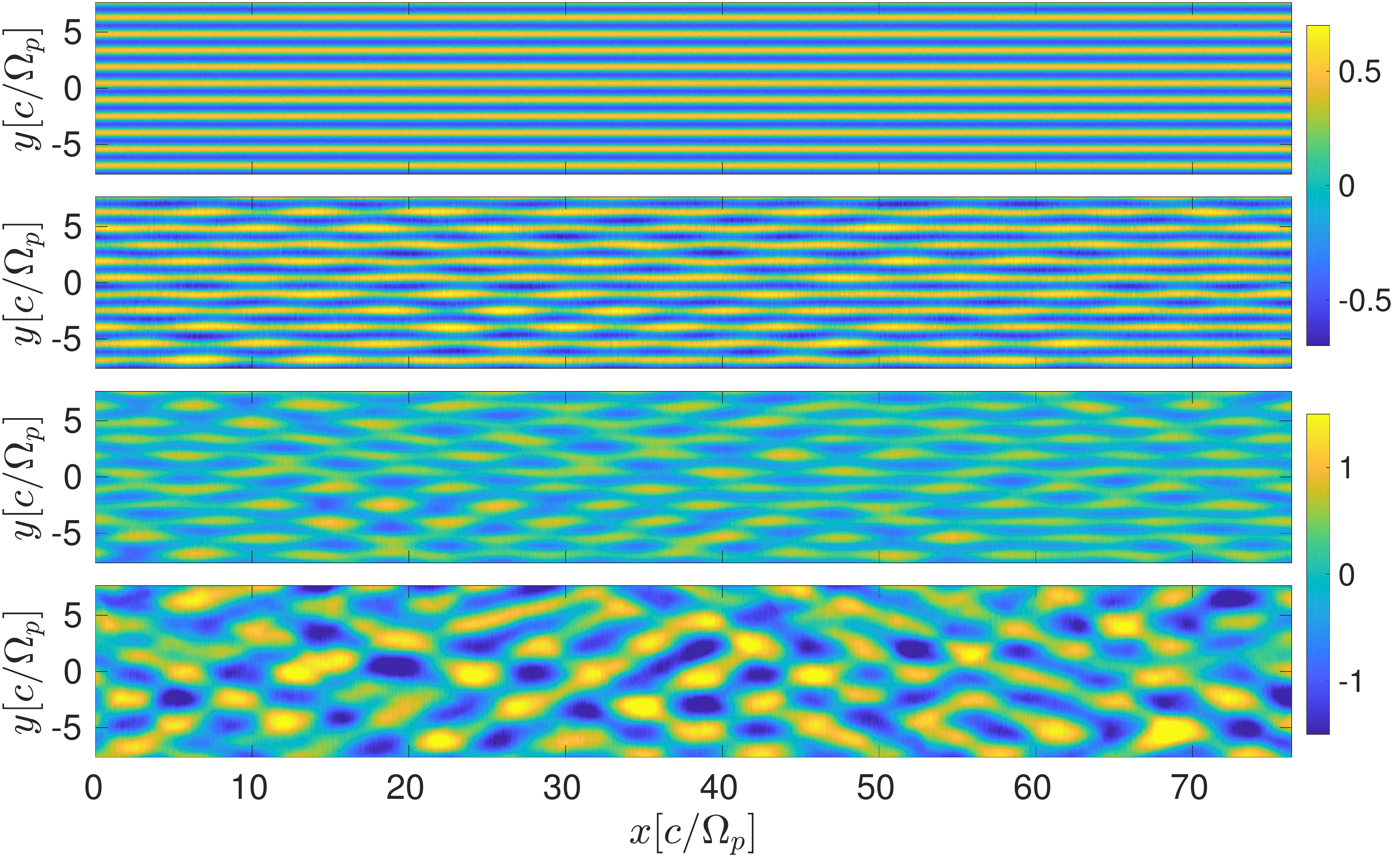}
  \caption{2D PIC simulation with the parameters of Fig.~\ref{fgr:floquet_asym_dom}: transverse magnetic field ($B_z$, normalized to $m\Omega_p/e$) at
  different times. From top to bottom: $t = 3,\ 32,\ 41,\ 58\ \Omega_p^{-1}$.}
  \label{fgr:Bz}
\end{figure*}

Figures~\ref{fgr:D1} and \ref{fgr:Bz} display the spatial structures of the cold-beam-positron density ($d_1$) and transverse magnetic field ($B_z$) at
different times. The patterns observed at $t=32\,\Omega_p^{-1}$ (\textit{i.e.}, in the linear phase of the primary instability) consists of an ensemble
of kink and bunching-type perturbations, as expected from the growth rate plateau of Fig.~\ref{fgr:disp_asym_fl}. At saturation ($t=41\,\Omega_p^{-1}$),
the original 1D filamentary structure has evolved into a quasiperiodic pattern of compressed positron bunches, aligned along two preferential directions,
tilted at angles $\simeq \pm 25^\circ$ relative to the flow axis. These angles are close to those characterizing
the dominant Floquet eigenmode ($\tan^{-1}\left( k_x/k_y \right) \simeq \pm 25.4^\circ$). As time increases,
these density islands tend to coalesce, giving rise to oblique modulations with increasing wavelength and tilted at larger angles ($\simeq \pm 35^\circ$, see
panels at $t = 58\,\Omega_p^{-1}$). Such unstable skew modes are reminiscent of the oblique instability arising in homogeneous asymmetric two-stream
systems \cite{Bret_2010a}.

\section{\label{sec:level1_disc}Conclusion}

To conclude, the stability of a periodic system of relativistic $e^-e^+$ current filaments, such as resulting from the current filamentation instability
(CFI), has been thoroughly examined in a warm-fluid framework. As the nonlinearity of the equilibrium filaments (quantified by the dimensionless parameter
$\xi_\alpha =\gamma_\alpha \left|\beta_\alpha\right| a_0/T_\alpha$) is increased, our numerical Floquet-type calculations predict a smooth transition from a dominant, purely
transverse, filament merging instability (FMI) to a relatively long-wavelength drift-kink instability (DKI). In a weakly nonlinear, symmetric configuration ($\xi_\alpha = \xi$),
the FMI is found to obey a dispersion relation close to that of the homogeneous CFI. As $\xi$ is raised, the FMI is progressively mitigated, and is eventually
overtaken by DKI modes when $\xi\gtrsim 2.5$. In the strongly nonlinear limit, the filaments are well described by the relativistic Harris solution and behave essentially
independently from one another. As a result, the dominant DKI modes share the periodicity of the unperturbed system, and their properties are similar to
those analytically derived for a single isolated filament. We have briefly studied the stability of an asymmetric configuration composed of a hot beam
streaming against a cold background plasma. The dominant primary mode consists of a combination of kink and bunching-type perturbations. A PIC
simulation shows that this mode nonlinearly evolves into two oblique strings of high-density islands. Their subsequent decay is followed by the development
of a larger-scale oblique instability, similar to that arising in homogeneous multi-stream plasmas. This scenario may be relevant to the precursor region of
Weibel-mediated relativistic shocks \cite{Keshet_2008} where similarly asymmetric two-beam interactions are expected to arise.

For all the studied cases, the theoretical predictions are consistent with the results of PIC simulations.
Our study, however, was limited to 1D and 2D systems. Extending it to a 3D geometry is not straightforward, both as regards the construction of the initial, 2D periodic array of nonlinear filaments and the numerical resolution of the 3D eigenmodes, to which the technique developed in the present work is ill-suited. Finally, we have restricted our analysis to relativistic pair plasmas: its generalization to electron-ion plasmas will be the subject of a forthcoming study. 

\section*{\label{sec:Acknowledgements}Acknowledgements}

The authors would like to thank Guy Pelletier for fruitful discussions and comments. A.V. acknowledges financial support from the ILP LABEX
(under reference ANR-10-LABX-63) as part of the Idex SUPER, and which  is  financed  by  French  state  funds  managed  by  the ANR  within  the 
Investissements  d'Avenir  program  under reference ANR-11-IDEX-0004-02. This work was granted access to the HPC resources of TGCC/CCRT
under the allocation 2017-A0030407666 made by GENCI. M.L. has been financially supported by the ANR-14-CE33-0019 MACH project.

\appendix

\section{\label{app:fluid_equations}Linearized relativistic four-fluid equations}

In this section, we  present the full set of fluid equations used to obtain the Floquet-type eigenmodes of a periodic chain of current filaments.
We consider a 2D $(x,y)$ system made up of four plasma fluids (two electron species and two positron species), initially flowing in the $x$ direction.
The unperturbed system is a stationary, $y$-dependent solution to the fluid-Maxwell equations. In the linearized version of Eqs.~\eqref{eq:continuity}-\eqref{eq:Gauss},
one can therefore use the simplifications $\partial_t \to -i\omega$ and $\partial_x \to ik_x$. The perturbed quantities are defined such that
$p_\alpha = p_{\alpha 0}\left[f_\alpha (y) + \delta P_\alpha \right]$, $d_\alpha = d_{\alpha 0}\left[f_\alpha (y) + \delta D_\alpha \right]$,
$\mathbf{v}_\alpha = \beta_{\alpha 0} c \left[ \epsilon_\alpha  \mathbf{1}_x + \delta \mathbf{V}_\alpha \right]$, $\mathbf{E} = E_{0y} f_E(y) \bm{1}_y + B_{0z} \delta \mathbf{e}$
and $B_z = B_{0z} \left[ f_B(y)  + \delta b_z \right]$. For each species $\alpha$, we have introduced $\beta_{\alpha 0}$ the absolute value of its
unperturbed normalized velocity, $\epsilon_\alpha = \{ 1,-1,-1,1\}$ its drifting direction. The equilibrium density ($f_\alpha$), magnetic ($f_B$) and electric
($f_E$) profiles are computed numerically from Eqs.~\eqref{eq:density}-\eqref{eq:phi_stat}.
The linearized equations read
\begin{widetext}
  \begin{flalign}
  \partial_y \delta e_x &= - \frac{i\omega}{c} \delta b_z + i k_x  \delta e_y \,,  \label{eq:de_x} \\
  \partial_y \delta b_z &= -\frac{i \omega}{c} \delta e_x+ \frac{4 \pi}{B_{0z}} \sum \limits_\alpha q_\alpha \beta_{\alpha 0} d_{\alpha 0}
  \left[f_\alpha (y) \delta V_{\alpha x} + \epsilon_\alpha \delta D_\alpha \right] \,, \label{eq:db_z}  \\
  \partial_y \delta P_\alpha &= \frac{q_\alpha d_{\alpha 0} B_{0z}}{p_{\alpha 0}} \left\{ -\epsilon_\alpha f_B (y) \beta_{\alpha 0} \delta D_\alpha
  + f_\alpha (y) \delta e_y + \frac{E_{0y}}{B_{0z}} f_E(y) \delta D_\alpha + f_\alpha(y) \beta_{\alpha 0} \left[- \epsilon_\alpha \delta b_z - f_B(y) \delta V_{\alpha x} \right] \right\}
  \, \nonumber  \\
  &- \frac{\gamma_\alpha^2}{c^2}  \frac{1}{p_{\alpha 0}} \left(p_{\alpha 0} \frac{\Gamma_{\rm ad,\alpha}}{\Gamma_{\rm ad,\alpha} - 1} + 
  \frac{d_{\alpha 0}}{\gamma_\alpha} mc^2 \right) f_\alpha (y) \beta_\alpha c (-i \omega \delta V_{\alpha y} + 
  \epsilon_\alpha \beta_\alpha c\ i k_x \delta V_{\alpha y} ) \,, \label{eq:dP} \\
  \partial_y \delta V_{\alpha y} &=  \frac{i \omega}{c \beta_\alpha f_\alpha (y)}  \delta D_\alpha - i k_x \delta V_{\alpha x} - \frac{\partial_y f_\alpha (y)}{f_\alpha (y)} \delta V_{\alpha y} - 
  \frac{i k_x \epsilon_\alpha}{f_\alpha (y)}  \delta D_\alpha \, , \label{eq:dV_y}
  \end{flalign}
\end{widetext}
where the perturbed quantities $\delta e_y$, $\delta V_{\alpha x}$ and $\delta D_\alpha $ are given by
\begin{widetext}
\begin{align}
  \delta e_y &= - \frac{i}{\omega}  \left[ \frac{4 \pi c}{B_{0z}} \sum\limits_\alpha q_\alpha \beta_{\alpha 0}  d_{\alpha 0} f_\alpha(y) \delta V_{\alpha y}
  +i k_x c  \delta b_z \right] \,,  \\
  \delta V_{\alpha x} &=  \frac{1}{\frac{\gamma_\alpha^2}{c^2} (p_{\alpha 0} \frac{\Gamma_{\rm ad, \alpha}}{\Gamma_{\rm ad, \alpha} - 1} + \frac{d_{\alpha 0}}{\gamma_\alpha} mc^2)
  f_\alpha (y) \beta_{\alpha 0} c \left(-i \omega  + i \epsilon_\alpha \beta_{\alpha 0} c k_x  \right)} 
  \Biggl\{ - i k_x p_{\alpha 0}  \delta P_\alpha +  q_\alpha d_{\alpha 0} B_{0z} \Bigl[f_\alpha(y) \delta e_x  \Bigr. \Biggr.  \nonumber \\
  & \Biggl. \Bigl. + \beta_{\alpha 0} f_\alpha (y)  f_B(y) \delta V_{y,\alpha} \Bigl]- \Bigl[q_\alpha \beta_{\alpha 0}^2 d_{\alpha 0} B_{0z}  f_\alpha(y) \delta e_x
  + \epsilon_\alpha q_\alpha \beta_{\alpha 0}^2 d_{\alpha 0}   E_{0y} f_\alpha (y)  f_E(y) \delta V_{\alpha y} - i \epsilon_\alpha \omega c^{-1} p_{\alpha 0}  \beta_{\alpha 0} \delta P_\alpha \Bigr] \Biggr\}  \,, \\
  \delta D_\alpha &=  \frac{1}{\Gamma_{\rm ad, \alpha}} \left[ \delta P_\alpha 
  +  \epsilon_\alpha \Gamma_{\rm ad, \alpha}  f_\alpha (y) \frac{\beta_{\alpha 0}^2}{1 - \beta_{\alpha 0}^2} \delta V_{\alpha x} \right] \,.
\end{align}
\end{widetext}  
We can avoid the numerical estimation of $\partial_y f_\alpha$ in Eq.~\eqref{eq:dV_y} through the momentum equation in stationary state:
\begin{equation}
  \frac{\partial_y f_\alpha (y)}{f_\alpha (y)} = q_\alpha  \frac{d_{\alpha 0}}{p_{\alpha 0}} \left[E_{0y} f_E(y) - \beta_{\alpha 0 x}B_{0z}f_B(y)\right] \,.
\end{equation}
The system is solved using the Floquet theory for different points $(\omega,k_x) \in \mathbb{C}\times \mathbb{R}^+$, giving the corresponding characteristic
Floquet exponent $k_y$. We are interested in temporally unstable solutions, and hence we only retain the solutions with $k_y \in \mathbb{R}$.

\section{\label{app:DKI}Analytic solution of the relativistic drift kink instability}

In the strongly nonlinear (pinched) regime, each current filament can be described by an isolated system of two counterstreaming electron and positron fluids.
In the following we derive an approximate analytic solution to the relativistic drift-kink instability (DKI), valid in the vicinity of the filament center.
Our calculation is restricted to the case of two symmetric fluids. The positrons (resp. electrons) are taken to have a positive (resp. negative) velocity.

The set of equations is composed of the momentum, continuity and Maxwell's equations, closed by an adiabatic equation of state:
\begin{align}
  &\gamma^2_\alpha \left(p_\alpha + \epsilon_\alpha \right) \left(\partial_{ct} + \bm{\beta}_\alpha \cdot \bm{\nabla}\right) \bm{\beta}_\alpha
  =-\bm{\nabla} p_\alpha \nonumber \\
  &+ q_\alpha d_\alpha  \left(\mathbf{E} + \bm{\beta}_\alpha \times \mathbf{B}\right)
  - \bm{\beta}_\alpha \left( q_\alpha  d_\alpha \mathbf{E} \cdot \bm{\beta}_\alpha + \partial_{ct} p_\alpha \right) \,, \\
  &\bm{\nabla} \times \mathbf{B} = 4 \pi e \bm{\beta} \left(d_{+} + d_{-} \right) +  \partial_{ct} \mathbf{E} \,, \label{eq:max_amper_B2}\\
  &\bm{\nabla} \cdot \mathbf{E} = 4 \pi  e \left(d_{+} - d_{-} \right) \,, \\
  &\bm{\nabla} \times \mathbf{E} = - \partial_{ct}  \mathbf{B} \,, \\
  & \delta \left(\frac{p_{\alpha} \gamma^{\Gamma_{\rm ad}}}{d^{\Gamma_{\rm ad}}_\alpha} \right) = 0 \,,
\end{align}
where $e$ is the elementary charge, $\bm{\beta}$ is the drifting velocity vector of the positron, $\beta = \vert \bm{\beta} \vert $,
$\Gamma_{\rm ad}$ is the polytropic index and $\alpha = \left( +,- \right)$ refers to positrons and electrons.
We expand all fluid quantities to first order (for a generic variable $b\,=\, b_0 + \delta{b}$), and we introduce the following quantities:
\begin{flalign}
  \delta{b}^+ &= \delta{b}_+ + \delta{b}_- \,, \\
  \delta{b}^- &= \delta{b}_+ - \delta{b}_- \,.
\end{flalign}
For the sake of tractability, we make use of two assumptions. Firstly, we assume a purely transverse velocity perturbation,
\begin{equation}
\bm{\beta}_0 \cdot \delta \bm{\beta} = 0 \,.
\end{equation}
Secondly, we assume that the pressure balance condition in the filament is fulfilled:
\begin{equation}
\frac{\mathbf{B}_0 \cdot \delta{\mathbf{B}}}{4 \pi} + \delta p^+ = 0 \,.
\end{equation}
Using these two relations, the sum and difference of the linearized momentum equations take the following form
\begin{widetext}
\begin{align}
  &M_0 \left(-i\omega \delta{\bm{\beta}}^+ + \beta_0 i k_x \delta \bm{\beta}^- \right) = -\nabla \left(\delta p^+ \right)
  - e d_0 \mathbf{B}_0 \times \delta \bm{\beta}^- + 2e d_0 \bm{\beta}_0 \times \delta \mathbf{B} - e \mathbf{B}_0 \times \bm{\beta}_0 \delta d^+
  + i \omega \bm{\beta}_0 \delta p^- \,, \label{eq:pert_mom+} \\ 
    &M_0 \left( -i\omega \delta \bm{\beta}^- + \beta_0  i k_x \delta \bm{\beta}^+ \right) = -\nabla \left( \delta p^- \right) + 2e  d_0 \delta \mathbf{E}
  - e d_0 \mathbf{B}_0 \times \delta \bm{\beta}^+ - e \mathbf{B}_0 \times \bm{\beta}_0 \delta d^- - 2e \bm{\beta}_0 d_0 \bm{\beta}_0 \cdot \delta \mathbf{E}
  + i \omega  \bm{\beta}_0 \delta p^+  \,, \label{eq:pert_mom-}
\end{align}
\end{widetext}
where
\begin{equation}
  M_0 = \gamma_0^2 \left( \frac{\Gamma_\mathrm{ad}}{\Gamma_\mathrm{ad} - 1} p_0 + n_0 m c^2 \right) \, .
\end{equation}
Projecting Eqs.~\eqref{eq:pert_mom+} and \eqref{eq:pert_mom-} along the $x$ axis gives
\begin{equation}
  -i k_x \delta p^\pm + i \omega\ \delta p^\mp + e d_0 B_0 \delta \beta_y^\mp = 0 \,.
\end{equation}
Moreover, the linearized continuity equation writes
\begin{equation}
  - i \omega\ \delta d^\pm + i k_x \delta d^\mp + \partial_y d_0 \delta \beta^\pm_y + d_0 \partial_y \delta \beta^\pm_y = 0 \,.
\end{equation}
The adiabatic closure condition relates the perturbed pressure and density as
\begin{equation}
  \delta p^\pm = \Gamma_\mathrm{ad}  \frac{p_0}{d_0} \delta d^\pm \,.
\end{equation}
Combining the previous equations leads to the differential equation fulfilled by $\delta{\beta}^\pm_y$:
\begin{equation} \label{eq:DKI_diff}
  \left(\Gamma_\mathrm{ad} \partial_y \log d_0 + e d_0 B_0 \right) \delta \beta^\pm_y + \Gamma_\mathrm{ad} p_0 \partial_y \delta \beta^\pm_y = 0 \,.
\end{equation}
We assume that the current filament initially obeys the Harris solution,  
\begin{align}
  d_0 &= \gamma_0 n_0 \cosh^{-2} \left( y/l \right) \, , \\ 
  B_0 &= \sqrt{16 \pi n_0 T_0} \tanh \left(y/l \right) \,, 
\end{align}
with the characteristic width [Eq. \eqref{eq:Harris_width}]
\begin{equation}
  l = \frac{\sqrt{T_0}}{\gamma_0} \frac{c}{\Omega_p} \, .
\end{equation}
In the Harris equilibrium, the solution to Eq.~\eqref{eq:DKI_diff} is
\begin{equation}
\delta{\beta}^\pm_y = \delta{\beta}^\pm_y\left(0\right) \cosh\left(y/l\right)^{2 - 2/\Gamma_{\rm ad}} \,.
\end{equation}
This solution is even in $y$, consistent with the DKI. The validity of this expression is limited to the inner region of the current filament.

Linearizing the cross product of $\bm{B}$ with Eq.~\eqref{eq:max_amper_B2}, one obtains
\begin{align}\label{eq:pert_AM}
&- e \mathbf{B}_0 \times \bm{\beta_0} \delta{d}^+ + 2e d_0 \bm{\beta}_0 \times \delta \mathbf{B} + e d_0 \delta \bm{\beta}^- \times \mathbf{B}_0 \nonumber \\
&= -\nabla \left( \frac{\mathbf{B}_0 \cdot \delta \mathbf{B}}{4\pi} \right)
- \frac{i \omega}{4 \pi}  \mathbf{B}_0 \times \delta \mathbf{E} \,.
\end{align}
The drift term $\frac{i \omega}{4 \pi} \mathbf{B}_0 \times \delta \mathbf{E}$ is found to be negligible. Substituting Eq.~\eqref{eq:pert_AM} into
Eq.~\eqref{eq:pert_mom+} gives
\begin{equation}\label{eq:pert_mom}
  M_0 \left(-i\omega \delta{\bm{\beta}}^+ + \beta_0 i k_x \delta{\bm{\beta}}^- \right) =  i \omega \bm{\beta}_0 \delta{p}^-  \, .
\end{equation}
Therefore, $\delta{p}^-$ is of the same order as the $\delta \beta^\pm_x$ terms, and hence can be neglected (in agreement with the numerical
Floquet solutions). The projection of the two latter equations along the $y$-axis leads to the following system
\begin{flalign}
  M_0 \left(-i\omega \delta \beta^+_y  + \beta_0 i k_x \delta \beta^-_y \right) &=  0 \,,  \label{eq:pert_mom++} \\
  M _0\left(-i\omega \delta \beta^-_y  + \beta_0 i k_x \delta \beta^+_y \right) &=   2 e d_0 \delta E_y \,. \label{eq:pert_mom--}
\end{flalign}
The perturbed transverse electric field, $\delta E_y$, is related to $\delta \beta^\pm_y$ through
\begin{equation}\label{eq:E_y}
  \delta E_y = \frac{4 \pi}{i \omega B_0}  \left( e B_0  \delta \beta^-_y  - i k_x \delta p^+ \right)\,.
\end{equation}
Combining \eqref{eq:pert_mom++}-\eqref{eq:E_y} and evaluating the resulting system at the filament center finally gives
\begin{align}
  &\frac{\omega^2}{k_x^2} \left\{ 2 + \left[ \frac{k_x^2}{\Omega_p^2} - \frac{\omega^2}{\Omega_p^2} \right]
  - \Gamma_{\rm ad} \left[ 2 + \left( \frac{k_x^2}{\Omega_p^2} - \frac{\omega^2}{\Omega_p^2} \right)   \left(1+T \right)\right] \right\}  \nonumber \\
  & + 2 - \left[ \frac{k_x^2}{\Omega_p^2} - \frac{\omega^2}{\Omega_p^2} \right] - \Gamma_{\rm ad} \left[ 2 - \left( \frac{k_x^2}{\Omega_p^2}
  - \frac{\omega^2}{\Omega_p^2}\right) \left(1+T \right)\right] = 0\,.
\end{align}
The dispersion relation of the DKI therefore writes
\begin{equation}
  h_0 \frac{\Gamma^2}{\Omega_p^2} =  \sqrt{1+4 h_0 \frac{k_x^2}{\Omega_p^2}} - 1 - h_0 \frac{k_x^2}{\Omega_p^2} \,,
\end{equation}
where
\begin{equation} \label{eq:h}
  h_0 = 1 + \frac{\Gamma_\mathrm{ad}}{\Gamma_\mathrm{ad} - 1} T_0 \,.
\end{equation}
The growth rate reaches a maximum value of
\begin{equation}
  \Gamma_\mathrm{DKI,max} = \frac{1}{2}\frac{\Omega_p}{\sqrt{h_0}}
\end{equation} 
for $k_x=(\sqrt{3}/2)\Omega_p/\sqrt{h_0}$.

\section{\label{app:CFI}Analytic solution of the current filamentation instability}

The dispersion relation of the CFI in a system composed of two symmetric, relativistic pair plasma flows is obtained by linearizing
Eqs.~\eqref{eq:continuity}-\eqref{eq:Gauss} in the limit of vanishing fields. This yields the following linear system
\begin{widetext}
\begin{equation}
	\begin{pmatrix}
	-i \tilde{k}_y & \tilde{\Gamma}  & 0 & 0 & 0 & 0 \\
	\tilde{\Gamma} +\frac{2}{h_0 \tilde{\Gamma}} & -i \tilde{k}_y & \frac{\beta_0 \gamma_0 (h_0-\Gamma_{\rm ad} T_0)}{\Gamma_{\rm ad} h_0}
	& -\frac{\beta_0 \gamma_0 (h_0-\Gamma_{\rm ad} T_0)}{\Gamma_{\rm ad} h_0} & 0 & 0 \\
	0 & -\frac{\beta_0 \gamma_0}{T_0} & -i \tilde{k}_y & 0 & -\frac{\beta_0 \gamma_0^2 \left(h_0 \tilde{\Gamma}^2+1\right)}{T \tilde{\Gamma}} & -\frac{\beta_0 \gamma_0^2}{T_0 \tilde{\Gamma}} \\
	0 & \frac{\beta_0 \gamma_0}{T_0} & 0 & -i \tilde{k}_y & -\frac{\beta_0 \gamma_0^2}{T_0 \tilde{\Gamma}} & -\frac{\beta_0 \gamma_0^2 \left(h_0 \tilde{\Gamma}^2+1\right)}{T_0 \tilde{\Gamma}} \\
	-\frac{1}{\gamma_0 h_0} & 0 & \frac{\beta_0 T_0 \tilde{\Gamma} }{h_0}-\frac{\tilde{\Gamma}}{\beta_0 \Gamma_{\rm ad}} & 0 & -i \tilde{k}_y & 0 \\
	\frac{1}{\tilde{\Gamma} h_0} & 0 & 0 & \frac{\beta_0 T_0 \tilde{\Gamma} }{h_0}-\frac{\tilde{\Gamma}}{\beta_0 \Gamma_{\rm ad}} & 0 & -i \tilde{k}_y \\
	\end{pmatrix}
	\cdot
	\begin{pmatrix}
	\delta e_x \\ 
	\delta b_z \\ 
	\delta p_+ \\ 
	\delta p_- \\ 
	\delta v_{y+} \\ 
	\delta v_{y-}
	\end{pmatrix} 
	=0 \,,
\end{equation}
\end{widetext}
where we have introduced 
\begin{align}
  &\omega_p  = \Omega_p\,e^{-\gamma_0 a_0/2T_0} \,, \\
  &\tilde{\Gamma} = \Gamma/\omega_p \,, \\
  &\tilde{k}_y = k_y c/\omega_p  \,,
\end{align} 
and $h_0$ is given by Eq.~\eqref{eq:h}. To leading order in $1/\gamma_0^2$ , the dispersion relation writes
\begin{widetext}
  \begin{equation}
  h_0 \tilde{\Gamma}^2 = \frac{1}{2} \left(\sqrt{1+16/K_y^2} -1 \right) K_y^2
  - \frac{1}{2 \gamma_0^2 \left(h_0 - \Gamma_\mathrm{ad} T_0 \right)} \left[4 h_0 +  \Gamma_\mathrm{ad} T_0 K_y^2 + \frac{4 h_0
  +\left(8 + K_y^2\right) \Gamma_\mathrm{ad} T_0}{\sqrt{1+16/K_y^2}}\right] \,,
\end{equation}
\end{widetext}
with $K_y = \sqrt{h_0} \tilde{k}_y$.
In the ultrarelativistic case ($\gamma_0 \to \infty$), the maximum growth rate is
\begin{equation}
  \Gamma_\mathrm{CFI, max} = \frac{2}{\sqrt{h}} \omega_p =  \frac{2 }{\sqrt{h}}e^{-\gamma_0 a_0/2T_0} \Omega_p \,.
\end{equation}

\bibliographystyle{unsrt}
\bibliography{Bib}

\end{document}